\documentclass[aps,longbibliography]{revtex4-2}

\usepackage{graphicx,color,xcolor}
\usepackage{natbib}
\usepackage{color}
\usepackage{soul}
\usepackage{amsmath}
\usepackage{psfrag}
\usepackage{empheq}
\usepackage{bm}
\usepackage[normalem]{ulem}
\usepackage{amssymb,amsmath}
\usepackage{dcolumn}
\usepackage{mathtools}
\usepackage{subcaption}
\usepackage{array}
\usepackage[color=blue!20!white,textsize=tiny,textwidth=0.8in]{todonotes}
\usepackage{multirow}
\usepackage{enumitem}

\usepackage{pgfplots}
\pgfplotsset{compat=newest}
\usetikzlibrary{plotmarks}
\usetikzlibrary{arrows.meta}
\tikzset{>={Stealth[scale=1]}} 
\usetikzlibrary{calc,patterns,angles,quotes}
\usepgfplotslibrary{patchplots}
\usepackage{grffile}
\usepackage{amsmath}

\newcommand{\BibitemShut}[1]{}

\newsavebox{\astrutbox}
\sbox{\astrutbox}{\rule[-5pt]{0pt}{20pt}}

\definecolor{case1}{rgb}{0,0,0.6}
\definecolor{case2}{rgb}{0,0.6,0}
\definecolor{case3}{rgb}{0.6,0,0}
\definecolor{case4}{rgb}{0.6,0.6,0}
\definecolor{case5}{rgb}{0.6,0,0.6}

\begin{document}

\title{Identification of triadic phase coupling in wall-bounded turbulence\\using the bispectrum}

\author{Clayton P. Byers}
\email[Corresponding author: ]{clayton.byers@trincoll.edu}
\affiliation{Department of Engineering, Trinity College, Hartford, CT 06106, USA}
\author{Subrahmanyam  Duvvuri}
\affiliation{Department of Aerospace Engineering, Indian Institute of Science, Bengaluru 560012, India}

\date{xx; revised xx; accepted xx.}

	\begin{abstract}
    The direction and magnitude of energy transfer between turbulence scale brought about by external forcing on a turbulent boundary layer are uncovered through the bispectrum, bicoherence, and biphase. The bispectrum is a third-order, complex-valued spectrum of the streamwise velocity that preserves the phase information between triadically consistent scales. Normalized bispectrum is the bicoherence, a measure of the relative amount of energy at a higher frequency that results from quadratic phase coupling of two lower frequencies. The phase of the bispectrum, the biphase, measures the phase lag between the high frequency and two lower frequencies that add to it, unveiling whether a triadic interaction produces a forward or reverse cascade of energy. Summing the bispectrum over triadically consistent frequencies allows a spectral decomposition of the velocity skewness and asymmetry, unveiling the triadically active scales in the energy transfer processes. An average sense of energy transfer is inferred from the phase of this skewness spectrum, which shows that scales smaller than the boundary layer thickness contribute to a forward cascade on average, while those larger than the boundary layer thickness have a mix of forward and reverse events. These measures show that the forced scales in the perturbed boundary layer have a mixture of forward and reverse energy transfer processes for different sets of triadic scales and wall-normal locations, providing a method of quantifying the effects of external perturbations on turbulent flows without any need for artificial filtering.
	\end{abstract}

\maketitle

\section{Introduction}
The coupling between scales and the transfer of energy between them is an important area of study in wall turbulence. Starting with the demonstration of scale coupling between inner and outer regions of a turbulent boundary layer by \citet{rao1971bursting} and the study of phase relationship between scales by \citet{brown1977large}, many studies have followed that investigate the nature of large-scale motions and the interactions between large and small scales and the interscale energy transfer \cite[see for instance][and references therein]{jimenez2012cascades,baars2015wavelet,he2017space,wang2021coherent}. Large-scale motions have been found to modulate the small scales in turbulent boundary layers \cite{hutchins2007large,mathis2009large,chung2010large} and have been quantified with a correlation coefficient in \citet{mathis2009large}.  Using a temporal cross-correlation based on \citet{bandyopadhyay1984coupling}, \citet{jacobi2013phase} found that small-scale motions in the streamwise and wall-normal directions of motion lead the large-scale fluctuations in the streamwise direction. The organization of small scales has been identified to depend on height and amplitude, depending on whether the smaller scales are averaged with large-scale motions (LSMs) or very-large-scale motions (VLSMs) \cite{saxton2022amplitude}. These findings were consistent with previous studies on the spatial organization of vortical structures \cite{adrian2007hairpin} and the burst and sweep depiction of structures within the boundary layer \cite{kline1967structure}. The lead and lag relationship of the scales was further formalized through a phase calculation of the cross-correlation function, leading to the finding that the phase difference between large and small scales in a wall-bounded flow was negative and would pass through $-\pi/2$ at the critical layer, becoming more negative further from the wall \cite{jacobi2021interactions}. \citet{jacobi2021interactions} then demonstrated that the VLSMs in a flow can be represented by a single scale and that the phase relationship between the streamwise small-scales and VLSM is predictable.

External perturbation of distinct scales within a turbulent flow can not only disrupt the energy cascade but also lead to modulation in energy content across the energy spectrum. There has been ample research into wall actuation for drag reduction in turbulent flow applications \cite{quadrio2011drag,yagiz2012drag,zhang2020active,ricco2021review}, including a recent study that investigated the physical mechanisms which allow spanwise forcing at the wall in a turbulent boundary layer to result in a reduction of skin friction drag \cite{marusic2021energy}. Two different forcing mechanisms were utilized; oscillations at frequencies comparable to small eddies in the near-wall region resulted in a significant drag reduction but at an energy cost exceeding the savings, but oscillations at frequencies comparable to large-scale eddies further from the surface produced a drag reduction at considerably less power consumption \cite{marusic2021energy}. \citet{deshpande2023relationship} found that the drag reduction is associated with enhanced coupling between the inner and outer scales, where wall actuation causes all scales to be closer in phase. This phase behavior was tied to the amplitude modulation coefficient of \citet{mathis2009large} by invoking the relationship among triadically consistent modes from \citet{duvvuri2015triadic}. The resulting phase of the cospectrum between outer scales and the ``outer scale envelope'' of the inner scales was found to go from out-of-phase towards in-phase in both actuating cases \cite{deshpande2023relationship}. Furthermore, any non-canonical effect such as wall roughness, pressure gradients, or wall-forcing in a turbulent flow will enhance triadic interactions \cite{lozier2024revisiting}. The phase relationship among all scales within the turbulent flow is intimately tied to the energy cascade and non-Gaussian behavior. \citet{wang2024role} found that freezing the phase relationship among scales in 3D isotropic turbulence results in an attenuated forward cascade, and reversing the phase field can generate an inverse cascade. Therefore, insight into the phase relationship among scales within a turbulent flow field is essential to understand the relationship between external perturbations and the resulting flow statistics.

The present effort is motivated by the experimental study of \citet{duvvuri2016nonlinear}, in which a turbulent boundary layer is externally forced with two distinct frequencies. The experimental data of \citet{duvvuri2016nonlinear} is subject to further analysis by a spectral tool that does not require any artificial filtering of scales. The goal is to develop both a formal tie between the phase-relationship among scales in the flow field and draw out the modifications to the energy content and cascade induced by the external forcing.

\section{The Bispectrum and Bicoherence} \label{section:analysis}

A bispectral analysis identifies the triadic interactions of coupled frequencies such that frequencies $f_1$ and $f_2$ interact to directly influence $f_3 = f_1+f_2$. Bispectrum and bicoherance (a normalization of the bispectrum) have been used in a wide range of disciplines,  analyzing biomedically relevant signals such as EEG signals in rats during various vigilance states \cite{ning1989bispectral}, determining autism spectrum disorder diagnostics from EEG signals \cite{pham2020autism}, investigating the growth of modes that lead to transition in a hypersonic boundary layer \cite{chokani2005nonlinear}, determining the correlation between forced oscillation modes in wind power production \cite{lyu2020correlation}, and assessing the interaction of dynamic modes in the boundary layer of a sandstorm \cite{liu2022evolution}. Time series that exhibit a higher level of noise may conceal dynamic states that are not immediately apparent in the power spectrum, but can be extracted with a bicoherence calculation \cite{george2017detecting}. As was done in a DNS of a channel flow by \citet{cui2021biphase}, the bicoherence extracted the relationship between large and small scales and their triadic interactions.  In this study, the streamwise velocity in both canonical and periodically perturbed turbulent boundary layers are utilized in the calculation of the bispectrum and bicoherence to highlight this coupling and extract relevant information that the power spectrum alone cannot reveal. 

A detailed review of the theory of the bispectrum $B(f_1,f_2)$ is given by \citet{sigl1994introduction} and is formally written as
\begin{equation}
     B(f_1,f_2) = \langle X_i(f_1)X_i(f_2)X^*_i(f_1+f_2) \rangle, \label{eq:bis}
\end{equation}
where $i$ refers to an epoch of time from the time series, a total of $L$ epochs are extracted from the data set, $X_i(f)$ is the Fourier transform of the $i^{th}$ epoch of the time series $x(t)$, and $X^*_i(f_1+f_2)$ is the complex conjugate of $X_i(f_1+f_2)$. Note that $ \langle \cdot \rangle = \frac{1}{L}\sum_{i=1}^{L} (\cdot)$ represents ensemble averaging. The bispectrum produces a non-zero magnitude associated with pairs of frequencies $f_1$ and $f_2$ that have a consistent phase relationship with frequency $f_1+f_2$ \cite{matsuoka1984phase,sigl1994introduction}. To illustrate, imagine two separate signals $a(t)$ and $b(t)$ comprising the following frequency, amplitude, and phase content:
\begin{subequations}
  \begin{align}
    a(t) &= \sin(2\pi f_1t+\phi_1)+\sin(2\pi f_2t+\phi_2)+\sin\left(2\pi (f_1+f_2)t+(\phi_1+\phi_2)\right)+\sin\left(2\pi (f_1-f_2)t+(\phi_1-\phi_2)\right), \label{eq:ex1} \\
    b(t) &= \sin(2\pi f_1t+\phi_\alpha)+\sin(2\pi f_2t+\phi_\beta)+\sin\left(2\pi (f_1+f_2)t+\phi_\gamma\right)+\sin\left(2\pi (f_1-f_2)t+\phi_\delta\right),\label{eq:ex2}
  \end{align}
\end{subequations}
where the frequencies $f_1$ and $f_2$ and the phases $\phi_1$ and $\phi_2$ are independent, and the phases $\phi_\alpha$, $\phi_\beta$, $\phi_\gamma$, $\phi_\delta$ are random and independent. Equations \ref{eq:ex1} \& \ref{eq:ex2}  will have the same power spectrum, as all phase information within the signal is ignored in the calculation. Therefore, any information about quadratic phase coupling within the signal cannot be unveiled through the power spectrum alone and requires the bispectrum. If triadically consistent frequencies have a persistent phase relationship across the individual epochs (i.e., equation \ref{eq:ex1}), then the bispectrum will produce distinct peaks at those frequencies of interest. An analogous expression can be formed in analyzing wavevectors $\vec{k}_3= \vec{k}_1+\vec{k}_2$ rather than frequency. 

Much like the power spectrum $P(f) = \langle|X(f)|^2\rangle$, the magnitude of $B(f_1,f_2)$ represents an ``energy'' at given frequency pairs in the signal, but does not directly reveal the nature of the triadic interaction of frequencies or any information about the direction of energy tranfer in the non-linear process that gives rise to quadratically phase coupled triads. In particular, it is possible for three frequencies $f_1$, $f_2$, and $f_3 = f_1+f_2$ to exist without a physical mechanism that resulted in their phase coherence. In this instance, the bispectrum would still produce a non-zero result. In order to assess the level of phase coherence between two frequencies, the bispectrum is normalized as:
\begin{equation}
    b(f_1,f_2) = \frac{|B(f_1,f_2)|}{\sqrt{P(f_1)P(f_2)P(f_1+f_2)}}. \label{eq:bic}
\end{equation}
Note that there are multiple normalization schemes, and the term ``bicoherence'' is sometimes misleading as the normalization in Equation \ref{eq:bic} is not bound by $0\leq b(f_1,f_2)\leq 1$, but instead is more of a ``skewness spectrum''  \cite{HINICH2005Normalizing, fackrell1995quadratic,fackrell1996bispectral}. To clearly see what this measure provides, the velocity signal $u(t)$ is written as a sum of Fourier modes: 
\begin{equation}
    u(t) = \alpha_1\sin(\omega_1t+\phi_1)+ \alpha_2\sin(\omega_2t+\phi_2)+\cdots+\alpha_i\sin(\omega_it+\phi_i)+\cdots+ \text{to $\omega_\infty$}, \label{eq:velocity modes}
\end{equation} 
where $\omega = 2\pi f$ is the angular frequency, $\alpha$ and $\phi$ are the magnitude and phase angle of a given frequency mode, respectively, and $0<\omega_1<\omega_2<\cdots<\omega_i<\cdots<\omega_\infty$. Therefore, the fourier transform of $u(t)$ will yield a series of modes $\hat{u}_n(\omega_n)$ that each contain an amplitude and phase. For a multiplicative interaction over which the nonlinearity acts on two modes $\hat{u}_1$ and $\hat{u}_2$ and results in quadratic coupling, and assuming uniform random distributions of phases exist for each given frequency, the nonlinearity can be written as:
\begin{equation}
    \alpha_1\sin(\omega_1t+\phi_1)*\alpha_2\sin(\omega_2 t+\phi_2) \Rightarrow \alpha_c \sin\left[(\omega_1+\omega_2)(t-\tau)+\phi_1+\phi_2\right]+\alpha_c \sin\left[(\omega_1-\omega_2)(t-\tau)+\phi_1-\phi_2\right], \label{eq:nonlinear}
\end{equation}
where $\alpha_c$ is the amplitude that results from the quadratic phase coupling and $\tau$ is a time delay over which the multiplicative nonlinearity occurs \cite{duvvuri2015triadic,cui2021biphase}. Note that this delay of $\tau$ can depend on the combinations of frequencies and therefore is not expected to take on the same value throughout the spectral domain, even for the sum or difference modes of the same frequency pair, or even for the same frequency pair at different locations within the boundary layer \cite{jamvsek2003time,shils1996bispectral}. By defining a phase delay $\phi_D = (\omega_1+\omega_2)\tau$ and letting $\omega_1+\omega_2 = \omega_3$, then the Fourier mode $\hat{u}(\omega)$ at $\omega_3$ can be represented as:
\begin{equation}
    \hat{u}(\omega_3) = \alpha_3 e^{i\phi_3}+\alpha_c e^{i(\phi_1+\phi_2-\phi_D)}, \label{eq:phase delay}
\end{equation}
where $\alpha_3$ and $\phi_3$ are associated with the uncoupled energy that exists at that frequency without any contribution from the nonlinear interaction. Given these definitions, the bicoherence $b(f_1,f_2)$, where $f_1+f_2 = f_3$ for the phase coupling, can be approximated in the ensemble limit as \cite{sigl1994introduction,cui2021biphase}: 
\begin{equation}
    b(f_1,f_2)\approx \frac{\alpha_c}{\sqrt{\alpha_c^2+\alpha_3^2}}. \label{eq:bic meaning}
\end{equation}
This normalization provides a clear way to interpret the fraction of energy at a given frequency due to the quadratic phase coupling from $f_1$ and $f_2$ \cite{cui2021biphase}. When no phase coupling occurs, $b(f_1,f_2)\to 0$, while strongly coupled signals approach $b = 1$. The bicoherence has an advantage over other detection methods for phase-amplitude coupling due to the direct calculation in Fourier space, whereas other methodologies rely on filtering that can introduce artificial constraints and be subject to bias \cite{kovach2018bispectrum,zandvoort2021defining}. Therefore, the bicoherence provides a direct look into the phase coupling without any arbitrary selection of filter parameters that can affect the results. 

The phase coupling highlighted from the bicoherence can be further understood by calculating the phase angle between frequency components. The biphase is calculated as the angle of the bispectrum in the complex plane, or
\begin{equation}
    \beta(f_1,f_2) = \tan^{-1}\left(\frac{\text{Im}\{B(f_1,f_2)\}}{\text{Re}\{B(f_1,f_2)\}}\right), \label{eq:biphase}
\end{equation}
where Im and Re represent the imaginary and real component of the bispectrum, respectively. By substituting Equations \ref{eq:velocity modes} and \ref{eq:phase delay} into \ref{eq:bis}, then using that result in Equation \ref{eq:biphase}, the biphase reduces, in the ensemble limit, to
\begin{equation}
    \beta(f_1,f_2)\approx \phi_D.
\end{equation}
This clearly shows the biphase is a measure of the phase delay in a quadratically coupled frequency pair, and moreover reveals the nature of the cascade of energy. As shown in \citet{cui2021biphase}, a temporally-based signal such as those collected from hot wires will have a phase of $(0,+\pi)$ for a reverse cascade of energy (from large to small scales), and $(-\pi,0)$ represents a forward cascade. This is evident from the phase delay $\phi_D$ representing the temporal lag in the nonlinear process, and therefore negative biphase implies the higher frequency $f_3$ has a phase that leads the larger scale motions, consistent with the evidence of a forward cascade as seen in previous work \cite{chung2010large,hutchins2011three,jacobi2013phase,baars2015wavelet,jacobi2021interactions, cui2021biphase}.

\section{Experimental Setup and Flow Statistics}
\begin{figure}
     \centering
    	\includegraphics[trim={1 220 1 210}, width=\textwidth, clip]{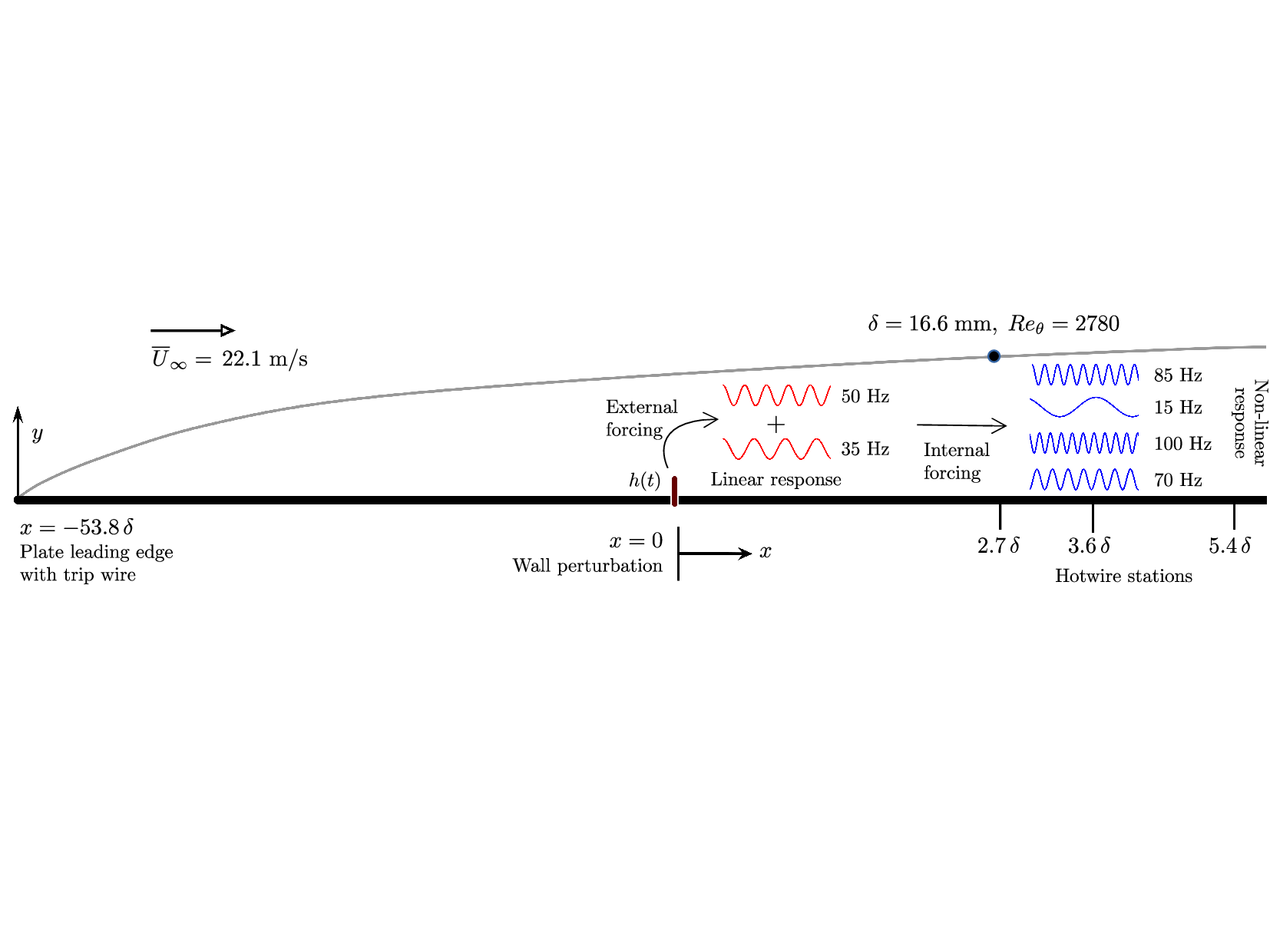}
    	\caption{Schematic of the experimental setup. Figure reproduced from \citet{duvvuri2016nonlinear}.}
        \label{fig:schem}
\end{figure}

The experimental setup has been extensively discussed in \citet{duvvuri2014phase,duvvuri2016nonlinear,duvvuri2017phase}, but is summarized here for completeness. A flat plate boundary layer with freestream velocity $\bar{U}_\infty=22.1$ m/s develops after being tripped with a wire at the leading edge. The instantaneous streamwise velocity is measured at a location where $\delta\approx 16.6$ mm and the momentum thickness Reynolds number is $Re_\theta \approx 2780$. Using the Coles-Fernholz empirical relationship, the friction Reynolds number is estimated at $Re_\tau \approx 940$ for the canonical flow \cite{duvvuri2015triadic}. In addition to the canonical flat plate boundary layer flow, two different external forcing conditions imposed by an actuator are introduced at a distance of $x/\delta = 2.7$ upstream of the measurement location. A single-frequency forcing condition takes place where the height $h(t)$ of the actuator is given as (in mm)
\begin{equation}
    h(t) = 0.4 + 0.4\cos(2\pi f_b t), \label{eq:forcing1}
\end{equation}
where $f_b = 50$ Hz is the frequency of oscillation. This results in the oscillation height extending to a maximum non-dimensional height of $h/\delta = 0.049$ and an RMS value of $h_{rms}/\delta = 0.033$. Additionally, a second forcing case with two forcing frequencies takes place with the wall perturbation given by
\begin{equation}
    h(t) = 0.4 \left[2+\cos\left(2\pi f_a t\right)+\cos\left(2\pi f_b t\right)\right]. \label{eq:forcing2}
\end{equation}
The perturbation has the parameters $f_a = 35$ Hz and $f_b =  50$ Hz. This gives a maximum non-dimensional height of $h/\delta = 0.096$ and $h_{rms}/\delta = 0.054$. In both cases, $h(t) = 0$ corresponds to the wall. A schematic of the experimental setup is shown in Fig.~\ref{fig:schem}.

\begin{figure}
     \centering
    	\includegraphics[width=\textwidth]{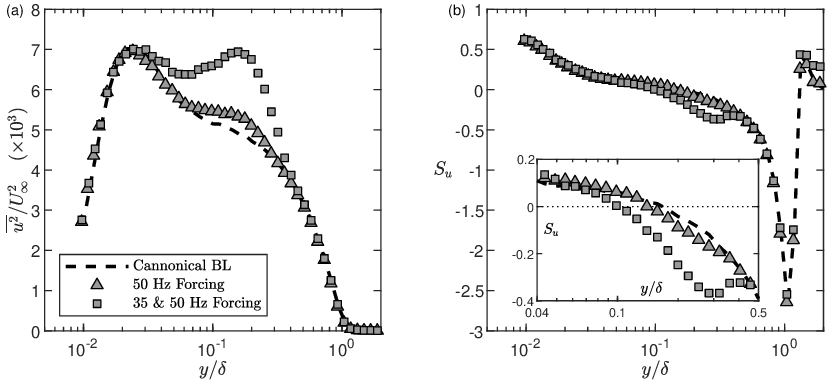}
    	\caption{Statistics of the three experimental cases. (a) The streamwise variance as a function of wall position, normalized in outer units: canonical flow (- -); single forcing frequency ($\triangle$); two forcing frequencies ($\square$). (b) The skewness ($S_u$) with the same symbols representing the three cases. Inset shows the region of maximal deviation from the canonical flow.}
        \label{fig:stats}
\end{figure}

The velocity statistics of the three different flow cases at $x = 2.7\delta$ are shown in Fig.~\ref{fig:stats}. Note that all subsequent analysis shown herein takes place at this downstream location. The streamwise variance (normalized by outer units) is shown in Fig.~\ref{fig:stats}(a), where it is apparent that the introduction of the perturbation at the wall results in an enhanced $\overline{u^2}$ throughout the region of $0.02 \leq y/\delta \leq 0.4$. The case of two forcing frequencies has an additional peak in the variance at $y/\delta = 0.2$, which is more than twice the distance from the wall that the actuator extends. This influence of the perturbations is reflected in the skewness ($S_u = \overline{u^3}/(\overline{u^2})^{3/2}$) profiles in Fig.~\ref{fig:stats}(b), where the perturbed profiles show a deviation from the canonical flow over a similar range. The shift in the skewness towards more negative values reflects the increase in negative streamwise fluctuations, likely through enhanced ejection events. In both the variance and skewness profiles, the forced cases closely resemble the canonical cases beyond $y/\delta >0.5$.

\begin{figure}
     \centering
    	\includegraphics[width=\textwidth]{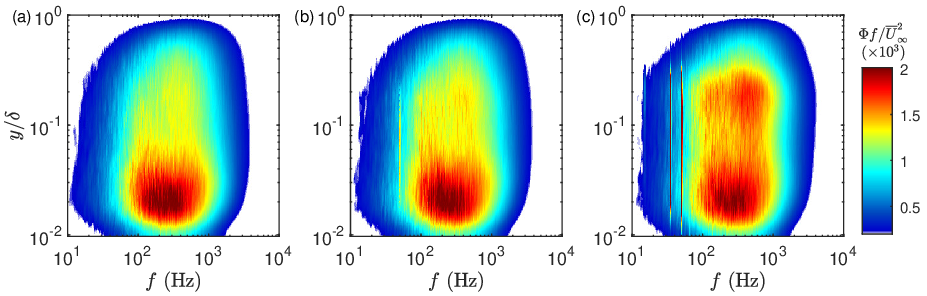}
    	\caption{Pre-multiplied velocity spectra for the three flow configurations, with the canonical (unforced) boundary layer in (a), the single forcing case in (b), and the two-frequency forcing in (c). }
        \label{fig:premult}
\end{figure}

This region of higher turbulence intensity is apparent in the spectrum, shown in the premultiplied form in Fig.~\ref{fig:premult}. The single-frequency forcing case shows enhanced energy at 50 Hz from a region of $0.02\leq y/\delta \leq 0.25$, and a broader region of increased energy between 100 and 1000 Hz in the region outside $y/\delta \geq 0.05$. Slightly visible in the power spectrum is the harmonic of 100 Hz, which has a slight increase in energy relative to its surroundings. Of interest are the resulting triadic interactions between the two-frequency forcing case seen throughout Fig.~\ref{fig:premult}(c). In addition to the prominent energy bands at the two forcing frequencies of $f_a=35$ Hz and $f_b = 50$ Hz, the triads that are combinations of sums, differences, and doubles of $f_a$ and $f_b$ are present throughout the majority of the boundary layer, with the prominence appearing to decay around $y/\delta \approx 0.4 \text{ to } 0.5$. Within the spectrum, the prominent triads appear to fade after $100$ Hz and do not appear distinct from the general spectral shape. The region of $0.05 \leq y/\delta \leq 0.3$ and $100 \leq f \leq 1000$ also has a significantly increased amount of energy compared to both the canonical and single forcing frequency, aligning with the variance plots in Fig.~\ref{fig:stats}.

\section{Results of the Triadic Interactions}

Using the bispectrum, bicoherence, and biphase, a new methodology for extracting information related to phase coupling of triad and the nature of energy transfer is developed here. Additionally, the relationship between the bispectrum and the velocity skewness will be explored, providing another look into how scales interact. Lastly, the streamwise spectral energy transfer function will be computed using the bispectrum. These different measures will be compared to each other in addition to previous methodologies utilized in earlier studies, noting that the bispectral measures are all a direct calculation from the Fourier transform, requiring no filtering or arbitrary selection of scale separation.

\subsection{The Bispectrum}

Notable with the magnitude of the bispectrum is that it does not contain information about phase coupling. Equation \ref{eq:bis} utilizes multiple epochs of the time series over which the data are averaged. If there is energy present at frequencies that have a mathematically triadic relationship, i.e.,~ $f_1+f_2 = f_3$ where all frequencies are strongly present in the spectrum, then $B(f_1,f_2)$ will have a larger magnitude compared to frequency pairs that do not have triadic relationships. However, these frequencies can be paired without any physical mechanism that couples them. Therefore, it must be noted that the calculation from Equation ~\ref{eq:bis} does not discriminate between physically coupled frequencies and triads that happen to exist without a mechanism driving their coupling. 

\begin{figure}
     \centering
    	\includegraphics[width=\textwidth]{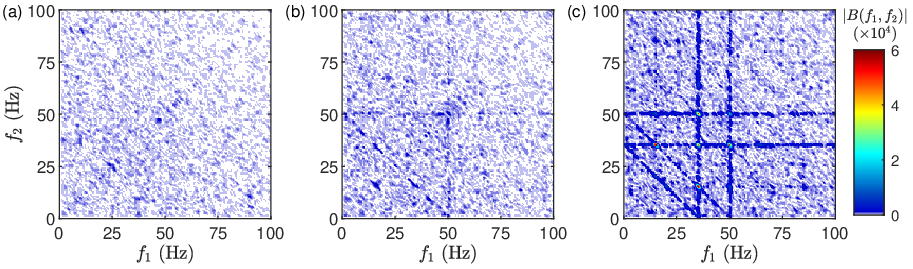}
    	\caption{Magnitude of the bispectrum at $y/\delta =0.2$ in the low frequency region for (a) the canonical boundary layer, (b) the single forcing frequency case, and (c) two-frequency forcing case.}
        \label{fig:bispec}
\end{figure}

Although the bispectrum may not give the level of coupling present, it does provide insight into which triads of frequencies are present in the flow field. Additionally, the symmetry of $B(f_1,f_2) = B(f_2,f_1)$ means that the data are equivalent across the reflection over the line of $f_1 = f_2$. The bispectrum is calculated at all wall-normal positions for all three flow conditions, while the $y/\delta = 0.20$ case is shown in Fig.~\ref{fig:bispec} as a representative example in the low-frequency region to highlight the triadic interactions from the forcing frequencies. This location is chosen because it corresponds to the peak in the variance of the two-frequency forcing case, as well as the region of enhanced energy content shown in Fig.~\ref{fig:premult}. 
Figure \ref{fig:bispec}(a) shows the canonical boundary layer, which does not show any particular pattern in the coupling of frequencies. However, the introduction of a single forcing frequency  produces a clear coupling as seen in Fig.~\ref{fig:bispec}(b), where the 50 Hz signal stands out across all frequencies. The strongest point occurs at $f_1 = f_2 = 50$ Hz, which clearly indicates that the harmonic triad of the forcing frequency coupling with itself is the most prominent interaction at this wall-normal location. Additionally, there is a diagonal intersecting the $f_1 = 50$ and $f_2 = 50$ Hz points, signaling that all low-frequency triads that add to 50 Hz have an enhanced bispectral signal. However, as will be seen in the bicoherence, this does not mean there is a significant energy transfer mechanism; $B(f_1, f_2)$ simply indicates that there is a prominence of energy at the two frequencies and their sum.

The triadic interactions become clear in Fig.~\ref{fig:bispec}(c), where the two-frequency forcing case shows the coupling of the forcing frequencies at 35 and 50 Hz, as well as a number of triads associated with these frequencies. The diagonal lines indicate that all lower-frequency pairs that add to either 35 or 50 Hz are prominent, similar to what was seen in Fig.~\ref{fig:bispec}(b) with 50 Hz. The horizontal and vertical lines are indications of a particular frequency (such as 35 Hz) and triads of it being present throughout the flow field and energy spectrum, where $35+f_2 = f_3$ are all present (and the equivalent for 50 Hz). The most prominent points in Fig.~\ref{fig:bispec}(c) correspond to the sum, difference, and harmonics of 35 and 50 Hz, as well as the triads of these triads.

\subsection{The Bicoherence and Biphase}
Calculating the bicoherence from Equation \ref{eq:bic} will highlight whether a triadic interaction found in the bispectrum is due to phase coupling. The denominator of Equation \ref{eq:bic} is analogous to the bispectrum of a signal in which all phase angles are zero, corresponding to maximal phase coupling \cite{sigl1994introduction}. The bispectral calculation will have magnitudes lower than the power spectral calculation when no phase coupling occurs, so the bicoherence is a relative measure of phase coupling, with $0$ corresponding to no phase coupling and increasing values corresponding to increased phase coupling. Note that this measure is not bound to an upper limit based on the choice of normalization, but this does not change the interpretation of these results. In addition, the statistical significance of the results can be determined. Although nonzero bicoherence values are associated with nonlinear interactions among frequency components, any finite-length time series will have nonzero bicoherence \cite{haubrich1965earth}. \citet{elgar1989statistics} determined the level of significance of bicoherence signals $b^2$ in different bin sizes $L$, providing a methodology to determine the statistical significance of a $b^2(f_1,f_2)$ signal from zero. The estimate of the variance of $b^2$, assuming a chi-square distribution near $b^2(f_1,f_2)=0$, yielded results that were nearly identical to the actual measured values \cite{kim1979digital,elgar1989statistics}. Assuming a significance level of $\alpha=0.05$ for the chi-square distribution, the values of $b^2$ that are not statistically significant compared to zero are rejected. For all data presented in this manuscript, this corresponds to $b^2<0.0226$. Therefore, the bicoherence plots below show $b(f_1,f_2) \geq 0.150$ as masked for statistical significance.

\begin{figure}
     \centering
    	\includegraphics[width=\textwidth]{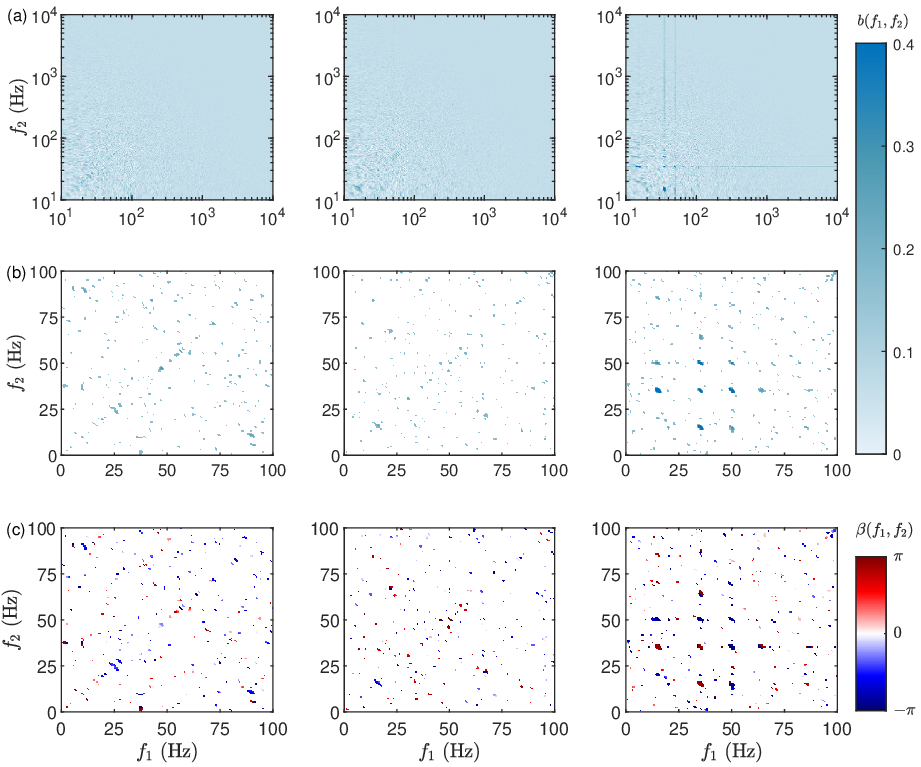}
    	\caption{All plots shown at $y/\delta =0.20$. (a) Bicoherence over the entire frequency domain (no statistical masking). (b) Bicoherence shown only in the low-frequency range. Data has been masked for statistical significance following \citet{elgar1989statistics}. (c) Biphase over the same frequency region in (b) for statistically significant values corresponding to the bicoherence. The left column is the canonical boundary layer, middle column is the single forcing frequency case, and the right column is the two-frequency forcing case.}
        \label{fig:bicohere}
\end{figure}

The bicoherence across all frequencies at a wall-normal location of $y/\delta = 0.20$ for the three forcing conditions is shown in Fig.~\ref{fig:bicohere} (a), with the canonical case in the left column, the single forcing frequency case in the middle column, and the two-forcing frequency case in the right column. The canonical and single forcing frequency cases both show a small amount of coupling across the entire domain but do not show any specific trend. In contrast, the two-frequency forcing case clearly shows lines of coherence at 35 and 50 Hz across all frequencies.  
Figure \ref{fig:bicohere}(b) shows the same calculation limited to frequencies below 100 Hz and masking for statistical significance, allowing the triadic interactions of the external forcing to stand out clearly. It is apparent that the canonical and single-forcing frequency cases do not have any particularly strong triads with the exception of self-coupling of frequencies. Note that the locations of statistically significant coupling in the canonical and single forcing frequency cases change with each $y/\delta$, which implies that there is no particular triadic interaction that persists throughout the flow (with the exception of $50 + 50 = 100$ Hz in the single forcing frequency case). In contrast to these two examples, the two-frequency forcing case shows strong coherence at combinations including 15, 20, 35, 50, 65, 70, and 85 Hz.  While the 15 and 85 Hz frequencies are triads formed by the difference and sum of the forcing frequencies, the 20 and 65 Hz frequencies are not algebraically related to any direct triad of 35 or 50 Hz. This indicates that in addition to the triads of the forcing frequencies being prominent in the boundary layer, secondary triadic interactions are also active. However, the strongest coupling appears at the sum, difference, and harmonic triads of 35 and 50 Hz, as seen in previous results \cite{duvvuri2016nonlinear,duvvuri2017phase}. It is important to note that there will always be some non-zero bicoherence among triads throughout the entirety of the boundary layer, and therefore other frequency pairs not corresponding to the forcing frequencies or their triads are not discussed.

Of note is that each wall-normal position will have different triads throughout the frequency domain that result in statistically significant levels of coherence, and the ones present throughout figure \ref{fig:bicohere}(b) that are not forcing frequencies do not necessarily appear as significant values at the wall-normal positions immediately above or below $y/\delta = 0.20$. Additionally, any example of a statistically significant triad that is not related to the forcing frequencies or their triads will have values of $b(f_1,f_2)$ that is smaller than the forced triads. This behavior persists over the majority of the near-wall region and out towards $y/\delta \approx 0.40$, which will be shown later. 

The nature of the energy transfer within these triads can be elucidated through the biphase, which is shown in Fig.~\ref{fig:bicohere}(c). \citet{cui2021biphase} show that the biphase gives the negative phase difference between large and small scales. Therefore, wherever a strong bicoherence is detected, the phase of the associated bispectrum yields the phase lag information between the interacting frequencies, with a phase of $0<\beta<\pi$ associated with the reverse cascade, and $-\pi < \beta < 0$ associated with the forward cascade \cite{cui2021biphase}. The canonical and single-frequency forcing cases show many triads with a forward cascade, while there are some combinations that appear to result in a reverse cascade. Note that the distribution of these points in the frequency plane will change at each wall-normal position. Instead, when looking at the two-frequency forcing case, there are distinct forward and reverse cascades for the particular triads of the forcing frequencies. It is clear that 50 Hz couples with 15, 35, and itself to have a forward cascade of energy to 65, 85, and 100 Hz, respectively. Likewise, 35 and 15 Hz form a triad to 50 Hz, but show a reverse cascade. This is due to the fact that the 15 Hz signal is arising from difference mode of 35 and 50 Hz. However, certain pairs, such as $35+35 = 70$ Hz, show a reverse cascade as well, which seems counterintuitive. Greater insight into these cascades can be found by looking at phase and coherence across the boundary layer by fixing one of the frequencies rather than fixing the wall-normal position \cite{jeffries1998experience, cui2021biphase}.

\subsection{Slices of the bicoherence and biphase}

\begin{figure}
     \centering
    	\includegraphics[width=\textwidth]{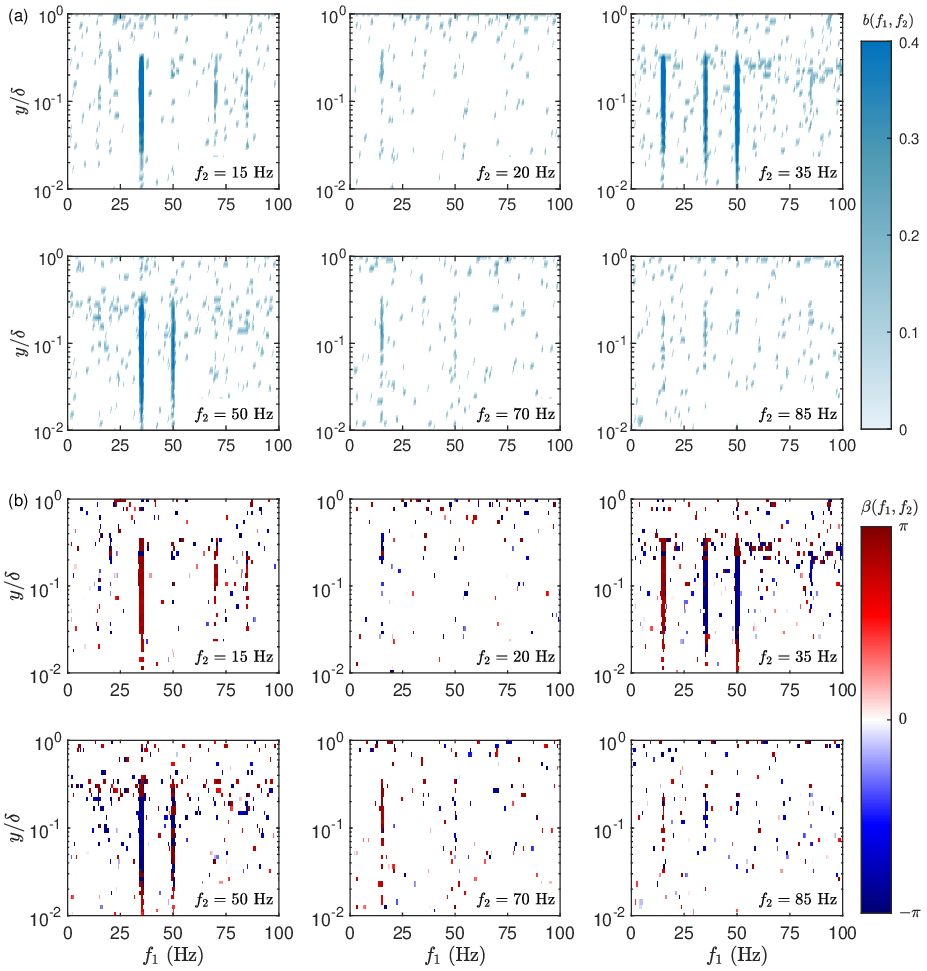}
    	\caption{Slices of the (a) bicoherence and (b) biphase at a set frequency $f_2$ with respect to coupling frequency $f_1$ and wall normal distance $y/\delta$. The six most prominent coupling frequencies of $f_1$ or $f_2$ are represented. All data has been masked for statistical significance.}
        \label{fig:slices}
\end{figure}

To better assess phase coupling throughout the boundary layer, ``slices'' of bicoherence and biphase will be taken at fixed frequency values for $f_2$ and plotted against all other frequencies $f_1$ across all wall-normal distances $y/\delta$. 
Figure \ref{fig:slices}(a) shows these slices of the bicoherence for 15, 20, 35, 50, 70, and 85 Hz, from top left to bottom right. 
Note that these plots are truncated to 100 Hz since the coupling beyond this frequency did not differ from the general background values. The strongest coupling is again seen with the main sum and difference modes of 35 and 50 Hz, and extends from the near-wall region to $y/\delta \approx 0.40$. The self-coupling harmonic triads of $35+35= 70$ Hz and $50+50 = 100$ Hz follow in magnitude of $b(f_1,f_2)$, just as seen in \citet{duvvuri2016nonlinear}, then further combinations secondary triads of frequencies that add to the original forcing frequencies or their immediate triads, such as $ 15 + 70 = 85$ Hz, $15 + 20 = 35$ Hz, $ 15 + 85 = 100$ Hz, and $ 35 + 85 = 120$ Hz. 

One particular characteristic of Fig.\ \ref{fig:slices} is how different triadic interactions have both a forward and a reverse cascade, dependent on the location within the boundary layer. The strongest coupling is shown between $50$ and $35$ Hz and has the greatest magnitude for $0.035<y/\delta<0.30$. This is expected as these are the two frequencies in which energy has been externally injected into the boundary layer, whereas other frequencies of interest are results of the triadic interactions in which energy must first cascade and transfer to that given scale. As Equation \ref{eq:bic meaning} demonstrates, the bicoherence is showing a relative energy that is present in the triad as a result of the quadratic phase coupling of the two lower frequencies. Therefore, this cascading effect across all triads will reduce the overall level of $b(f_1,f_2)$ as sums and differences of frequencies that are not 35 and 50 are considered. The phase relationship is shown in Fig.\ \ref{fig:bicohere}(b) and clearly shows the nature of the forward ($-\pi<\beta<0$) and reverse ($0<\beta<\pi$) cascade in the presence of external forcing. The interaction of 35 Hz with itself to produce 70 Hz, as well as 50 Hz with 35 Hz to produce 85 Hz, clearly shows a forward cascade throughout the majority of the interacting region as the biphase is negative. Conversely, the coupling between 15 Hz and 35 Hz has a strong reverse cascade, indicated by the positive phase. This demonstrates that energy is moving from the higher (forcing) frequencies of 35 and 50 Hz towards the lower frequency of 15 Hz. All three of these results are expected, as the forcing frequencies at 35 and 50 Hz will result in the interactions producing triadic sums to larger frequencies, indicative of a forward cascade of energy, while their differences that lead to smaller frequencies must be a reverse cascade. This was clearly shown in \citet{duvvuri2016nonlinear}, and is readily apparent in Fig.~\ref{fig:premult}, even if the nature of the forward and reverse cascades is not revealed by the spectral plot.

It is interesting to note that the interactions between 35 and 50 Hz, as well as their own harmonic interactions, all show a reverse cascade in some regions of Fig.\ \ref{fig:slices}(b). This would imply, for example, that the energy being sent to 85 Hz in the range of $0.03<y/\delta<0.30$ then returns in a reverse cascade through a triad with 50 Hz to excite the 35 Hz frequency in the near-wall region. While the triad of $35+50=85$ Hz is a natural result of the two forcing frequencies, the reverse cascade from 85 back to 35 and 50 Hz is only apparent through the use of the biphase. Additionally, other triads, such as $15+35 = 50$, $15+70 = 85$, and $15+85 = 100$, have strong reverse cascades, again implying that the higher frequencies are sending energy to excite the lower frequencies. 

Across all interactions in Fig.\ \ref{fig:slices}, strong coupling is shown throughout the region of $0.01<y/\delta<0.40$ for most triads involving 35 and 50 Hz, while less prominent triads have moderate coupling in limited regions of the boundary layer between $0.01<y/\delta<0.40$. Aside from the background levels of bicoherence that will be inherent with any finite time series of a nonlinear process \cite{haubrich1965earth}, there is no significant coupling from the dominant triads beyond $y/\delta>0.40$. For the 35 and 50 Hz data, there is broadband interaction with all frequencies present in the region of $0.1\leq y/\delta \leq 0.4$, which is the cause of the enhanced spectral energy present in Fig.\ \ref{fig:premult}(c). Extending beyond this low frequency regime, the plot of $b(f_1,f_2)$ involving the two forcing frequencies at $y/\delta=0.2$ in Fig.\ \ref{fig:bicohere}(a) clearly shows this coupling of the forcing frequencies across the entire frequency domain. This region of the boundary layer corresponds to the height in which the dominant synthetic modes investigated by \citet{duvvuri2017phase} had a maximal magnitude, followed by a significant reduction at $y/\delta \approx 0.4$. It also matches the results found through a phase-averaged assessment of the traveling waves produced by the triadic response modes \cite{duvvuri2016nonlinear}. Beyond $y/\delta = 0.4$, Fig.~\ref{fig:slices} show all frequencies in both the bicoherence and biphase reduce to background levels.

Figure \ref{fig:slices}(b) clearly shows how the forcing modes and their associated triads all have a reverse cascade in the near-wall ($y/\delta\leq 0.03$) region. Some triadic interactions are expected to be the result of a reverse cascade throughout the boundary layer, such as $50-35 = 15$ Hz, but the fact that the phase relationship is positive for all forcing modes and triads in this near-wall region is interesting to note. \citet{slomka2018nature} had shown that ``an inverse energy cascade may arise generically in the presence of flow-dependent narrow spectral forcing'', which is precisely the imposed boundary condition seen here. Furthermore, \cite{alexakis2018cascades} found that external mechanisms that break symmetry can result in split cascades, causing energy transfer to both small and large scales simultaneously. It is clear in Fig.~\ref{fig:slices}(b) that both inverse and forward cascades are arising from these external perturbations. Interestingly, all forcing frequencies and their triadic interactions have a near-wall reverse cascade. The location of the critical layer for the six most prominent bicoherence peaks associated with 50, 35, 85, 15, 100, and 70 Hz was found to be at $y/\delta \approx$ 0.07, 0.05, 0.06, 0.08, 0.075, and 0.045, respectively \cite{duvvuri2016nonlinear}. It is clear that this near-wall reverse cascade region lies below the associated critical layers of the forced modes and associated triads. Two potential mechanisms causing this could be a ``quasi-two-dimensional'' effect where these scales, being larger than the outer frequency scale of $f_o =U_\infty/2\pi\delta = 212$ Hz, are confined to more of a two-dimensional plane that is parallel to the surface and therefore experience that reverse cascade mechanism \cite{xiao2009physical}. Additionally, \citet{apostolidis2023turbulent} have found that inverse cascade events can be associated with the stretching of relative motions due to aligned fluctuation pairs. It is also possible that this inverse cascade in the near-wall region could be tied to an inner-outer scale coupling \cite{deshpande2023relationship}. However, this is speculative at this point and requires further investigation with multi-component velocity measurements and spatially resolved signals to tie the biphase with these physical phenomena.

A summary of the interactions and nature of energy transfer seen throughout Fig.\ \ref{fig:slices}, starting top left to bottom right for both (a) and (b), is given in Table \ref{tab:table1}.
\begin{table}[t]
\caption{\label{tab:table1} Summary of the predominant triadic interactions in the two-frequency forcing case as extracted from the bicoherence and biphase in Fig.~\ref{fig:slices}.}
\begin{ruledtabular}
\begin{tabular}{cccc}
\textrm{Triad (Hz)\footnote{These triads can be seen in multiple plots.}}&
\textrm{Coupling\footnote{As the magnitude of $b(f_1,f_2)$ varies, this is a rough qualitative summary.}}&
\textrm{Cascade}&
\textrm{Extent}\\
\colrule
$15+20 = 35$ & weak & forward \& reverse & $0.2 \leq y/\delta \leq 0.4$ \\
$15+35 = 50$ & strong & reverse & $0.01 \leq y/\delta \leq 0.4$ \\
$15+70 = 85$ & weak & reverse & $0.04 \leq y/\delta \leq 0.4$ \\
$15+85 = 100$ & weak & reverse & $0.04 \leq y/\delta \leq 0.2$ \\
$35+35 = 70$ & strong & forward & $0.02 \leq y/\delta \leq 0.4$ \\
$35+50 = 85$ & strong & forward & $0.03 \leq y/\delta \leq 0.2$ \\
 & strong & reverse & $0.01 \leq y/\delta \leq 0.03$ \\
 & strong & reverse & $0.2 \leq y/\delta \leq 0.4$ \\
$35+85 = 120$ & weak & forward & $0.1 \leq y/\delta \leq 0.3$ \\
$50+50 = 100$ & strong & forward \& reverse & $0.01 \leq y/\delta \leq 0.4$ \\
$50+85 = 135$ & weak & forward & $0.15 \leq y/\delta \leq 0.2$ \\
$70+50 = 120$ & weak & forward & $0.05 \leq y/\delta \leq 0.3$ \\
\end{tabular}
\end{ruledtabular}
\end{table}
In all cases, these results clearly indicate how the initial sum, difference, and harmonics (85, 15, 70, and 100 Hz, respectively) of the two forcing frequencies both send and receive energy from one another, and how these triads start to interact with each other as well to enhance energy content at frequencies that are not from the initial quadratic phase coupling (e.g. 20, 120, and 135 Hz). The bicoherence calculations for the triads beyond those listed in Table \ref{tab:table1} do not stand out relative to the background. Additionally, it becomes apparent that the quadratic phase coupling across the forcing frequencies and their related modes start to feed on each other, sending and receiving energy through multiple interactions. For instance, 15 Hz is not only the difference of the two forcing frequencies but also coupled with 70 and 85 Hz in a reverse cascade.

\subsection{Skewness and the Bispectrum} \label{section:skewness}
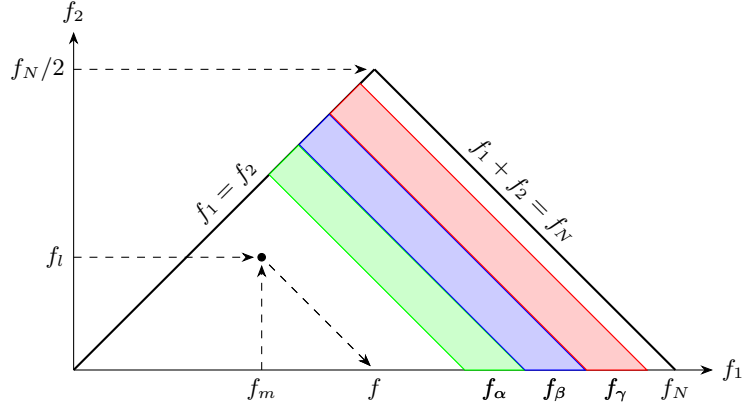
\begin{figure}
    \begin{tikzpicture}
        \draw[dashed, ->] (2.6, 1.4) -- (3.95, 0.05);
        \filldraw[fill=green!20, draw=green] 
        (2.6,2.6) -- (2.99,2.99) -- (5.99,0) -- (5.2,0) -- cycle;
        \node[below] at (5.6,0) {$f_\alpha$};
        \filldraw[fill=blue!20, draw=blue] 
        (3,3) -- (3.4,3.4) -- (6.8,0) -- (6,0) -- cycle;
        \node[below] at (6.4,0) {$f_\beta$};
        \filldraw[fill=red!20, draw=red] 
        (3.41,3.41) -- (3.81,3.81) -- (7.62,0) -- (6.82,0) -- cycle;
        \node[below] at (7.21,0) {$f_\gamma$};
        \draw[->] (0, 0) -- (8.5, 0) node[right] {$f_1$};
        \draw[->] (0, 0) -- (0, 4.5) node[above] {$f_2$};
        \draw[black,thick] (0, 0) -- (4, 4);
        \draw[black,thick] (4, 4) -- (8, 0);
        \node[above,rotate = 45] at (2.2,2.2) {$f_1 = f_2$};
        \node[above,rotate = -45] at (5.8,2.2) {$f_1+ f_2 = f_N$};
        \node[below] at (8,0) {$f_N$};
        \draw[dashed,->] (0,4) -- (3.9,4); 
        \node[left] at (0,4) {$f_N/2$};
        \node[left] at (0,1.5) {$f_l$};
        \draw[dashed, ->] (0, 1.5) -- (2.4, 1.5);
        \node[below] at (2.5,0) {$f_m$};
        \draw[dashed, ->] (2.5, 0) -- (2.5, 1.4);
        \node[below] at (4,0) {$f$};
        \draw[black,fill=black] (2.5,1.5) circle [radius=0.05];
        \draw[dashed, ->] (2.6, 1.4) -- (3.95, 0.05);
        \filldraw[fill=green!20, draw=green] 
        (2.6,2.6) -- (2.99,2.99) -- (5.99,0) -- (5.2,0) -- cycle;
        \node[below] at (5.6,0) {$f_\alpha$};
        \filldraw[fill=blue!20, draw=blue] 
        (3,3) -- (3.4,3.4) -- (6.8,0) -- (6,0) -- cycle;
        \node[below] at (6.4,0) {$f_\beta$};
        \filldraw[fill=red!20, draw=red] 
        (3.41,3.41) -- (3.81,3.81) -- (7.62,0) -- (6.82,0) -- cycle;
        \node[below] at (7.21,0) {$f_\gamma$};
        \end{tikzpicture}
    \caption{Domain of frequency pairs that add to the skewness and asymmetry spectrum. An example of a single triad $f_l+f_m=f$ is shown. Shaded regions show all triads adding to a frequency bin, with green for $f_\alpha$, blue for $f_\beta$, or red for $f_\gamma$.} \label{fig:decompose}
\end{figure}

The skewness $S_u$ of a stationary velocity signal can be shown to be a weighted sum of triads of Fourier components \cite{duvvuri2015triadic}. Since the bispectrum is a measure of the contribution to the third-order moment from frequency triads, $S_u$ can be constructed by summing over the different combinations of $f_1+f_2 = f_3$ throughout the spectral domain. Following \citet{elgar1987relationships}, the skewness can be directly related to bispectrum through the following summation:
\begin{equation}
    S_u+i A_u = \left[12\sum_l \sum_m B(f_l,f_m)+6\sum_n B(f_n,f_n)\right]\bigg/ (\overline{u^2})^{3/2}, \label{eq:SkewSum}
\end{equation}
where the lower symmetric region of the first quadrant of the $f_1$-$f_2$ plane is defined by $l<m$, $f_l+f_m < f_N$ for the Nyquist frequency $f_N$, $n < N/2$. The domain for this calculation is shown in Fig. \ref{fig:decompose}.  

From equation \ref{eq:SkewSum}, the real component, $S_u$, is the skewness of the velocity signal while the imaginary component, $A_u$, is an asymmetry measure of the waveform about a vertical axis \cite{elgar1987relationships}. 
This calculation for $S_u$ can be compared to the derivation of velocity skewness by \citet{duvvuri2015triadic}:
\begin{equation}
    S_u = \frac{6}{4(\overline{u^2})^{3/2}}\sum_{\substack{\forall\,l,m,n \\ \omega_l<\omega_m<\omega_n \\ \omega_l+\omega_m = \omega_n}} \alpha_l\alpha_m\alpha_n\sin(\phi_l+\phi_m-\phi_n)+\frac{3}{4(\overline{u^2})^{3/2}}\sum_{\substack{l=1 \\ \omega_n = 2\omega_l}}^\infty \alpha_l^2\alpha_n\sin(2\phi_l-\phi_n), \label{eq:skewness subbu}
\end{equation}
where the velocity signal is defined in the same fashion as in Equation \ref{eq:velocity modes}. It is apparent that adding the frequencies in the lower symmetric region in the first quadrant of the $f_1$-$f_2$ plane in which the bispectrum exists will cover only $1/8^{th}$ of the total plane. Therefore, multiplying Equation \ref{eq:skewness subbu} by a factor of 8 and performing the summation over the positive frequencies will result in the real component of Equation \ref{eq:SkewSum}. To understand the physical meaning of this, assume that frequency $f_n$ is due to quadratic phase coupling, therefore $\phi_n = \phi_l+\phi_m + \phi_D$ for some phase delay $\phi_D$ in the nonlinear interaction. It is clear that the skewness is then a measure of the total phase difference between triads of all frequencies. More particularly, the skewness shows the total contribution due to additive triadic interactions and harmonics and is thus a weighted sum of the phase difference between the lower and higher wavenumber modes. Therefore, the bispectrum at a given frequency pair is equivalent to a weighted measure of the phase difference between the lower and higher frequencies.

\begin{figure}
     \centering
    	\includegraphics[width=\textwidth]{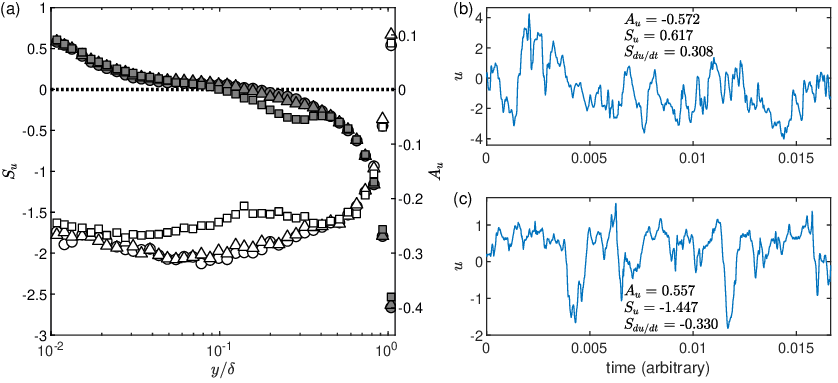}
    	\caption{(a) The real (left axes, solid symbols) and imaginary (right axes, hollow symbols) components from Equation \ref{eq:SkewSum}, with the canonical flow in circles, single forcing frequency case in triangles, and the two-frequency forcing case in squares. (b,c) Extracted waveforms of fluctuating velocity $u(t)$ at an arbitrary time to visualize the relationship between $A_u$, $S_u$, the skewness of the velocity derivative $S_{du/dt}$, and the waveform shape.}
        \label{fig:SkewnessCalc}
\end{figure}

The calculation for Equation \ref{eq:SkewSum} is shown in Fig.~\ref{fig:SkewnessCalc} for all three forcing conditions, noting that the resulting values of $S_u$ are identical to those shown in Fig.~\ref{fig:stats}(b) for all three cases. The asymmetry $A_u$ is negative across the entire boundary layer and nearly identical between the canonical and single forcing frequency cases, but moves towards zero in the two-frequency forcing case. Note that it can be shown that $A_u$ is identically equal to the skewness of the Hilbert transform of $u$ and is also related to the skewness of the slope of the timeseries of $u$ \cite{elgar1985observations}. To better understand this measure, Fig.~\ref{fig:SkewnessCalc}(b) and (c) show excerpts of a velocity signal at arbitrary times. The values of the asymmetry $A_u$, skewness $S_u$ and the skewness of the time derivative of the velocity $S_{du/dt}$ are inset in each plot. 
It is apparent in Fig.~\ref{fig:SkewnessCalc}(b) that this excerpt of the waveform has a generally negative slope, highlighted by the negative $A_u$. A positive value for the skewness of the velocity derivative indicates that it trends negative (the median value of $du/dt$ is lower than the mean), matching this measure of $A_u$. The opposite holds for Fig.~\ref{fig:SkewnessCalc}(c), where the generally positive waveform slope is characterized by a positive $A_u$ and negative derivative skewness. This asymmetry term has been used to relate the biphase and physical symmetry of surface waves \cite{elgar1985observations, xie2019effect} and to link the effects of a phase-lag to the transport of sediment under changing levels of wave asymmetry \cite{ruessink2009modeling}. The results shown in Fig.~\ref{fig:SkewnessCalc} indicate a negative asymmetry throughout the boundary layer, implying that the time series of $u$ tends towards a generally negative temporal derivative in its shape. The predominantly negative asymmetry is indicative of a general trend towards sudden sweeps followed by longer ejection periods, as more low-momentum fluid is being disrupted at the wall and moved out into the boundary layer. The addition of the perturbation at the wall will disrupt this action, creating enhanced mixing in a wake-like region behind the actuating rib. This manifests with the asymmetry increasing towards zero in the region where the forcing frequencies and associated triads are most active, meaning that the sweeps and ejections are more balanced. Although negative velocity skewness would generally indicate predominantly positive velocity fluctuations (seen in the outer boundary layer), the negative asymmetry indicates a general trend of a rapid increase (sweep) in velocity followed by a slowly decreasing velocity over longer times (ejection process). The two-frequency forcing case shows an increase of $A_u$ towards zero, indicating that the sweep and ejection events are moving towards being balanced.

\begin{figure}
     \centering
    	\includegraphics[width=\textwidth]{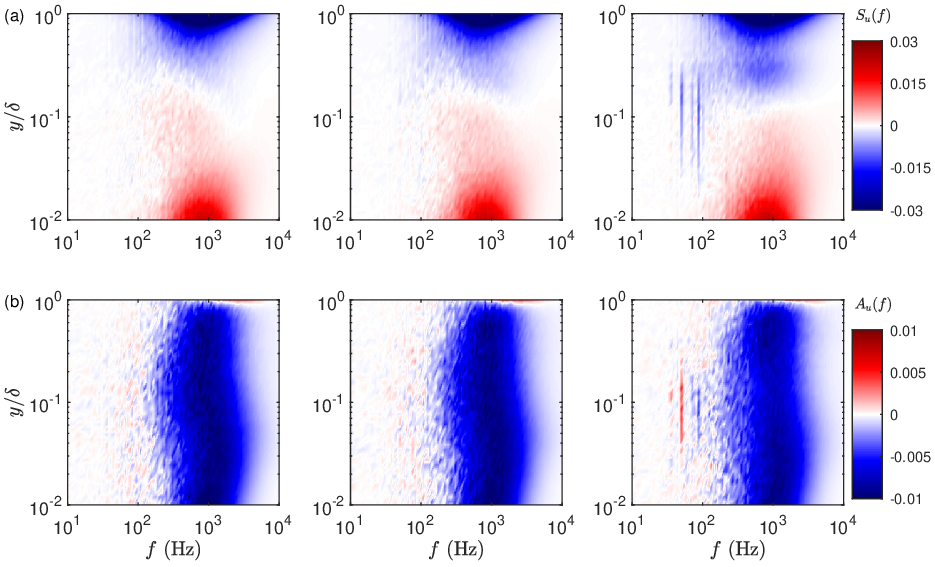}
    	\caption{Triadic summations of frequencies $f_1+f_2 = f$ that contribute to the skewness $S_u$ and asymmetry $A_u$ of Equation \ref{eq:SkewSum}. (a) Skewness contributions and (b) Asymmetry contributions. Canonical flow is shown on the left, single-frequency forcing in the middle, and two-frequency forcing on the right. Summations are combined over logarithmic frequency bins, highlighting the contribution from the high-frequency region.}
        \label{fig:SkewnessFreq}
\end{figure}

The velocity skewness and asymmetry can be decomposed spectrally by breaking the summations of Equation \ref{eq:SkewSum} into frequency bins such that $l$, $m$, and $n$ all fall within a range of frequencies $f$ consistent with $f_l<f_m<f_n$, with $f_l+f_m = f$ on the first summation, and $2f_n = f$ on the second summation. This will yield all triadic interactions of lower frequencies that add to the higher frequency $f$, providing the ``skewness and asymmetry spectrum'' as a function of the higher frequency involved in the triad \cite[see for instance][]{matsuoka1984phase,midya2022spectral}. Examples of these bins are given within the domain sketched in Fig.~\ref{fig:decompose}. By binning all triads of $f_l$ and $f_m$ that add to a single frequency $f$, the relative importance of all triadic interactions associated with frequencies lower than $f$, and therefore their contribution to the skewness of a signal, will be revealed \cite{midya2022spectral}. Note that this summation will be represented over logarithmically spaced bins on the frequency domain, therefore emphasizing the influence of the higher frequencies. As seen in Fig.~\ref{fig:bispec}, the magnitude of the bispectrum is largest for lower $f$ and decreases with increasing frequency, much like the power spectrum. Thus, the largest values of this spectral decomposition for a single frequency would occur at lower frequencies. However, the summations will be performed over logarithmically spaced frequency bins (see shaded regions in Fig.~\ref{fig:decompose}), skewing the overall appearance of the magnitude towards the higher end. For instance, the bins highlighted in Fig.~\ref{fig:decompose} could correspond to regions spanning a frequency range of 180 Hz in green, 200 Hz in blue, and 240 Hz in red to obtain the logarithmically spaced appearance. This resulting visualization will be similar to a pre-multiplied energy spectrum, as Fig.~\ref{fig:premult} does not indicate that the most energy is in the range of $100\leq f\leq 1000$, but instead emphasizes the concentration of energy in that region. This decomposition is shown in Fig.~\ref{fig:SkewnessFreq}(a) for all three cases for $S_u(f)$. 
It is clear that the greatest contribution to the skewness arises from the moderate frequency range of $100 \leq f \leq 2000$ Hz throughout the boundary layer for all three cases. In the canonical case, there is a zone between $0.1 \leq y/\delta \leq 0.3$ where the contributions to $S_u(f)$ shift from positive to negative, corresponding to the zero crossing of $S_u(f)$ in Fig.~\ref{fig:SkewnessCalc} in that same region. The single forcing case shows a slight downward shift in this zone that results in the skewness being slightly lower than the canonical case in the same region. For the two-frequency forcing case, the transition between positive and negative contributions moves closer to the wall, in addition to the triadic interactions prominently displaying in the region of $0.02 \leq y/\delta \leq 0.4$. Note that the triad of 85 Hz features most prominent, while the 35 Hz forcing frequency has a slight positive contribution close to the wall. All other triads manifest as a negative contribution. These results align with the statistics seen in Fig.~\ref{fig:stats}, where the greatest deviation from the canonical values is in the range of $0.02 \leq y/\delta \leq 0.4$, which is consistent with the regions of increased bicoherence and triadic interactions seen in Fig.~\ref{fig:slices}. It is important to note that this calculation of $S_u(f)$ is for all additive triads, so the nature of the reverse cascade of $50-35 = 15$ Hz is not captured in this skewness spectrum at 15 Hz, but instead arises as an enhanced magnitude at 50 Hz.

The spectral contribution to asymmetry $A_u(f)$ is shown in Fig.~\ref{fig:SkewnessFreq}(b) and shows a consistently negative value throughout the boundary layer and for all frequencies in the canonical case, with some positive values in the low-frequency region. The greatest magnitude occurs in the same region of $100 \leq f \leq 3000$ Hz as was seen for $S_u$. There appears to be minimal change between the canonical and single forcing frequency case, but the two-frequency forcing case shows a positive contribution from both the 35 and 50 Hz signals. In contrast, the triads of the two forcing frequencies provide a smaller negative contribution. This results in a general upward shift in $A_u$ as seen in Fig.~\ref{fig:SkewnessCalc}(a). Since the phase of the bispectrum gives the direction of energy transfer and $A_u(f)$ is from the imaginary component, the positive values associated with 35 and 50 Hz indicates that they are contributing to a reverse cascade, as was clear in Fig. \ref{fig:slices}(b). What Fig.~\ref{fig:SkewnessFreq} shows is the net effect of all triads that add to these frequencies, highlighting that most of the triads that add to 35 or 50 Hz are involved in a reverse cascade. In contrast, triads that add to values greater than these forced frequencies are predominantly involved in a forward cascade.

The spectral decomposition of $A_u$ may provide an indication of the scale of sweep or ejection events, which are closely tied to momentum transfer and are considered a fundamental part of the self-similar family of eddies that comprise the logarithmic layer in turbulence \cite{jimenez2012cascades}. Although some studies have assessed the imbalance in the stress contribution between sweeps and ejections through expansion models using third-order mixed moments of velocity fluctuations \cite{katul2006relative}, the spectral decomposition of skewness and asymmetry can potentially lend insight into the scales of motion that are tied to energy transfer as well as the sweeping or ejecting motion. The negative $A_u(f)$ throughout the higher frequencies is an indication that the waveforms associated with these triads tend toward strong spikes followed by longer decaying slopes in the time series of $u(t)$. Alternatively, the positive $A_u$ in the lower frequency regime is associated with gentler positive slopes followed by sharp declines in $u(t)$. In order to tie these measures with sweeps and ejections, this spectral decomposition needs to be performed with a dataset that has the wall-normal velocity as well, so further speculation is withheld in this discussion.

\begin{figure}
     \centering
    	\includegraphics[width=\textwidth]{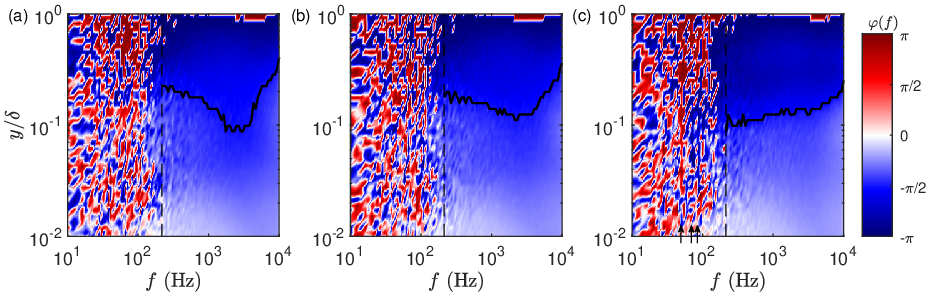}
    	\caption{The ``average phase'' $\varphi(f)$ from Equation \ref{eq:bulkphase} for the (a) canonical flow, (b) single-frequency forcing, and (c) two-frequency forcing cases. Black line indicates location of $\varphi = -\pi/2$. Vertical dashed line indicates the outer frequency scale $U_\infty/2\pi\delta = 212$ Hz. Locations of 50, 70, and 85 Hz are highlighted with arrow markers in the two-frequency forcing case.}
        \label{fig:bulkphase}
\end{figure}

Another measure of the scalar energy transfer process can be inferred by looking at the phase relationship of this spectral breakdown of skewness and asymmetry. An ``average phase'' is calculated by looking at the phase relationship of this spectral decomposition: 
\begin{equation}
    \varphi(f) = \tan^{-1}\left[\frac{A_u(f)}{S_u(f)} \right], \label{eq:bulkphase}
\end{equation}
which provides an aggregate of all triadic interactions of $f_1+f_2 = f$ by computing the average phase. This can be compared with the analysis of \citet{jacobi2021interactions}, where the amplitude modulation coefficient of \citet{mathis2009large} was shown to be zero when the phase relationship between large and small scales was $-\pi/2$. This implies the amplitude modulation coefficient is ``the cosine of the phase separating the scales,'' \cite{jacobi2021interactions}, while Equation \ref{eq:bulkphase} instead shows an aggregate phase of all triads adding to a particular frequency. Therefore, while previous studies were able to show how small scales physically lead an isolated large scale through these filtered measures, the calculation of $\varphi(f)$ will instead show the average lead/lag relationship of all triadic interactions that add to a given higher frequency. Note that this calculation will improve at higher frequencies as there are more triadic combinations over which the sum can be computed, and therefore the low-frequency regime may be noisy. The average phase is shown in Fig.~\ref{fig:bulkphase} for all three cases. 
The low frequency region has an immediately intriguing result in that the positive phases are almost exclusively below the outer frequency scale of $f_o =U_\infty/2\pi\delta = 212$ Hz, shown in the dashed line, in all three cases.  An exception exists in the very outer boundary layer where intermittency effects are likely scrambling the average phase. Note this scale matches with the filter cutoff used in the cross-correlation analysis between large and small scales \cite{mathis2009large,jacobi2013phase}. This result with all frequencies below the outer frequency scale seems to imply that the reverse cascade effects are confined (on average) to the largest of scales in the streamwise direction. Triadic interactions adding to a frequency below $f_o$ do not exhibit a clear trend with the exception of 50, 70, and 85 Hz in the two-frequency forcing case. The 50 Hz signal shows a strong positive phase, while 70 and 85 Hz show negative phase. There is also a small region of 100 Hz around $y/\delta = 0.1$ in the two-frequency forcing case that has a negative phase.  As was evidenced in both Fig.~\ref{fig:bicohere} and Fig.~\ref{fig:slices}, there are forward and reverse cascade events throughout the boundary layer for $f\leq 100$ Hz. As such, Fig.~\ref{fig:bulkphase} may indicate that the larger scales and their triads are in both a forward and reverse cascade process to continually redistribute energy among themselves.

Beyond $f_o$, Fig.~\ref{fig:bulkphase} clearly shows a negative average phase throughout the boundary layer for all frequencies under the three different flow conditions. This indicates that there is an average forward cascade of energy from all lower-frequency triads that adds to $f$. The phase is close to zero near the wall and decreases as $y/\delta \to 1$, approaching $-\pi$. This means the small scales start in-phase with the large scales close to the wall and increasingly lead the large scales away from the wall, consistent with the findings of prior studies \cite{mathis2009large,duvvuri2015triadic,jacobi2021interactions}. The line of $-\pi/2$ is indicated in all three cases with a solid black line and shows a trend of moving closer to the wall with added external forcing. In analyzing the correlation coefficient between a single large-scale mode and the remaining high-frequency content, \citet{jacobi2021interactions} found that the phase relationship was found to be $-\pi/2$ at the location where the correlation was zero and is in the vicinity of the critical layer for that scale. As Equation \ref{eq:bulkphase} and the results in Fig.~\ref{fig:bulkphase} are averages over all triadic interactions adding to that frequency, the comparison cannot be one-to-one. However, the general trend indicates the same phase relationship as seen in \citet{jacobi2021interactions} with respect to the relationship between large and small scales. The difference here is that Equation \ref{eq:bulkphase} provides a breakdown of this phase relationship across the spectral domain. In addition to $-\pi/2$ representing the location of a critical layer, this phase was shown to be the value that corresponds to maximum energy transfer though the partial interscale energy transfer mechanism from large to small scales \cite{cui2021biphase}. This location should also correspond to the outer spectral energy peak as noted by \cite{mathis2009large}. As evident in Fig.~\ref{fig:bulkphase}, the location of $\varphi=-\pi/2$ remains around $y/\delta = 0.1$ to 0.2, moving closer to the wall with the additional forcing. This corresponds well with the outer spectral energy peak in the pre-multiplied spectrum in Fig.~\ref{fig:premult}. Consistent with \citet{Li2023Quantifying}, these large-scale forcing frequencies result in increased coherence between large and small scales in the near-wall region, as evidenced by the location of $\varphi = -\pi/2$ shifting closer to the wall.

The most significant takeaway from this average phase calculation shown in Fig.~\ref{fig:bulkphase} is how information about critical layers, peak energy transfer, and scale alignment is all contained within this plot without any subjective filtering. Note that the filter frequency used to isolate the small scales corresponds to the outer frequency $f_o$ in \citet{jacobi2021interactions} and a similar scale within 20\% of $f_o$ in \citet{duvvuri2015triadic}. This means that Fig.~\ref{fig:bulkphase} is able to recover the same results as amplitude modulation coefficients and filtered scale interactions from direct spectral summations using the bispectrum.

\subsection{Spectral Energy Transfer and the Bispectrum}

While the real components of the bispectrum were summed to determine the skewness, a premultiplied sum of the imaginary components is used determine the spectral energy transfer function between streamwise modes and wavenumbers. This will provide additional insight into the nonlinear transfer of energy highlighted in the biphase slices of Fig.~\ref{fig:slices}(b). It can be shown that the spectral energy transfer function in the streamwise direction, $\hat{T}(k)$, can be expressed as \cite{Pope2000,cui2021biphase}
\begin{equation}
  \hat{T}(k)=-k\,\text{Im}\left\{ \quad\sum_{\mathclap{k_1+k_2 = k}}B(k_1,k_2)+ \sum_{\mathclap{k_1-k_2 = k}} B(k_1 - k_2,k_2) + \sum_{\mathclap{k_2 - k_1 = k}} B(k_2 - k_1,k_1) \right\}. \label{eq:SpecTransfer}
\end{equation}
The first summation in the brackets of Equation \ref{eq:SpecTransfer} corresponds to the energy transfer associated with larger scales sending energy to smaller scales, while the other two summations represent combinations of interscale transfer. The minus sign is due to the fact that positive phase is associated with a reverse cascade from temporally based signals \cite{cui2021biphase}, thus the negative imaginary component corresponds to the direction of spectral energy transfer. Also note that this is in terms of the local wavenumber $k$, which is computed through Taylor's hypothesis using the local mean velocity. The term $\hat{T}(k)$ (or the temporal equivalent $\hat{T}(f)$) gives an averaged sense of the energy transfer over all triads that sum to $k$ (or $f$). This Equation can be broken into two parts:
\begin{equation}
  \hat{T}_{S}(k)=-k\,\text{Im}\left\{ \quad\sum_{\mathclap{k_1+k_2 = k}}B(k_1,k_2)\right\} \quad\text{and}\quad \hat{T}_{D}(k)=-k\,\text{Im}\left\{ \quad\sum_{\mathclap{k_1-k_2 = k}} B(k_1 - k_2,k_2) + \sum_{\mathclap{k_2 - k_1 = k}} B(k_2 - k_1,k_1) \right\},
  \label{eq:SpecTransferParts}
\end{equation}
where $\hat{T}_{S}(k)$ represents the triadic transfer from two larger scales to a smaller scale, or the sum of wavenumbers, and $\hat{T}_{D}(k)$ represents transfer between mixed large and small scales, or the difference of wavenumbers.

\begin{figure}
     \centering
    	\includegraphics[width=\textwidth]{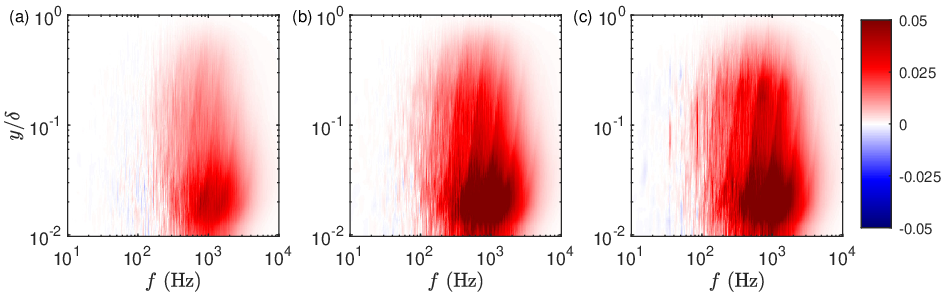}
    	\caption{Spectral transfer $\hat{T}(f)$ from Equation \ref{eq:SpecTransfer} for (a) the canonical flow case, (b) the one forcing frequency case, and (c) the two-frequency forcing case. }
        \label{fig:SpectralTransfer}
\end{figure}

Figure \ref{fig:SpectralTransfer} shows the streamwise spectral transfer function of Equation \ref{eq:SpecTransfer} for the three flow conditions, plotted with respect to frequency rather than wavenumber to remain consistent with previous plots. The two different forcing conditions show a clear enhancement of energy transfer in the higher frequency region compared to the unforced case. Figure \ref{fig:SpectralTransfer}(c) also shows strong forward energy transfer at 35 Hz in the $0.04 \leq y/\delta \leq 0.15$ range, and at 85 Hz out towards $y/\delta \approx 0.2$. Comparing Fig.~\ref{fig:SpectralTransfer} to Fig.~\ref{fig:bulkphase} shows that the strongest forward transfer occurs in the same frequency regions ($f>200$ Hz) that correspond to a negative ``average phase.''  This is apparent since the calculation of Equation \ref{eq:SpecTransfer} is a sum of all modes that couple to a triad at higher wavenumber, comparable to the summation for the skewness from Equation \ref{eq:SkewSum}. This result is consistent with the findings from \citet{lee2019spectral} for the streamwise component of energy transfer in a channel.

\begin{figure}
     \centering
    	\includegraphics[width=\textwidth]{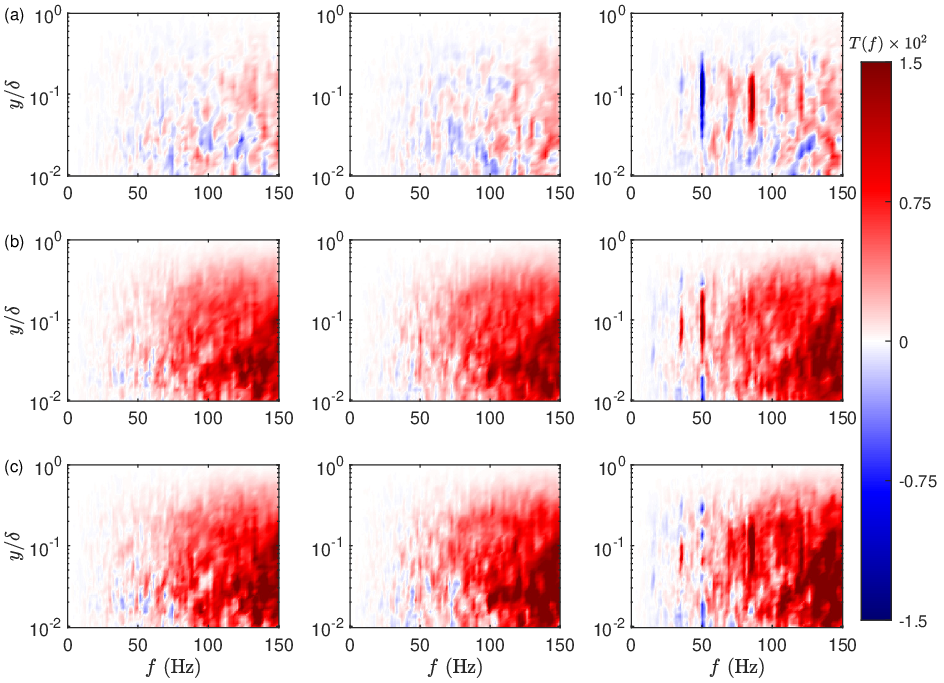}
    	\caption{Spectral transfer from Equations \ref{eq:SpecTransfer} and \ref{eq:SpecTransferParts} in the low frequency regime. (a) The summation contributions of triads $\hat{T}_{S}(f)$, (b) the difference contributions $\hat{T}_{D}(f)$, and (c) the total $\hat{T}(f)$ function. Left column is the canonical boundary layer, center column is the single forcing case, and the right column is the two-frequency forcing case. }
        \label{fig:SpecTransZoom}
\end{figure}

Greater detail on the influence of the forcing frequencies can be found by focusing on the $f\leq 150$ region for all three cases, shown in Fig.~\ref{fig:SpecTransZoom}, where the left column shows the canonical flow, the middle column shows the single-frequency forcing case, and the right column shows the two-frequency forcing case. Figure \ref{fig:SpecTransZoom}(a) shows the summation contributions of Equation \ref{eq:SpecTransferParts} which mimics the behavior of the ``average phase'' calculation in that there appears to be both forward and reverse energy transfer all throughout this low-frequency region. In all three flow configurations, the near-wall region shows more reverse cascade processes, comparable to what was seen in Fig.~\ref{fig:slices} with the positive phase of the strongly interacting modes for small $y/\delta$. The canonical and single-frequency forcing cases appear qualitatively similar, while the two-frequency forcing case shows a strong negative energy transfer for 50 Hz and a less intense negative transfer at 35 Hz. This is due to the interactions of lower frequencies (such as $15+20 = 35$ or $15+35 = 50$) requiring the reverse cascade to transfer energy to the lower frequency. Conversely, the prominent lines at 70, 85, and 120 Hz are all indications that their energy is due to lower frequency triads sending energy to these modes. The difference contributions $\hat{T}_D(f)$ in Fig.~\ref{fig:SpecTransZoom}(b) show a mostly positive energy transfer across the low-frequency regime in all cases. In the two-frequency forcing case, 35 and 50 Hz have positive energy transfer around the region of $0.03 \leq y/\delta \leq 0.2$, but reverse again closer to the wall and further away towards $y/\delta = 0.4$. Since the difference modes are associated with the displayed frequency being the smaller part of the triad (so $f_1+f = f_2$), then these reverse cascade regions match what was seen in the phase plots of Fig.~\ref{fig:slices}(b), in particular with the 15, 35, and 50 Hz coupling. The sum of Figs.~\ref{fig:slices}(a) and (b) gives the total interscale energy transfer $\hat{T}$ shown in Fig.~\ref{fig:slices}(c), which is the same information shown in Fig.~\ref{fig:SpectralTransfer} but focused on the low-frequency regime. Overall, it is clear that the in the two-frequency forcing case, there is significantly enhanced energy transfer to and from the forcing frequencies and the associated harmonics and triads compared to the canonical and single-frequency forcing cases. Additionally, while the bicoherence and biphase plots were able to show traids that were beyond the initial harmonics and sum/difference triads of 35 and 50 Hz, Fig.~\ref{fig:SpecTransZoom}(a) shows the first clear triadic interaction beyond 100 Hz (120 Hz being the result of 35 + 85 Hz). 

\begin{figure}
     \centering
    	\includegraphics[width=\textwidth]{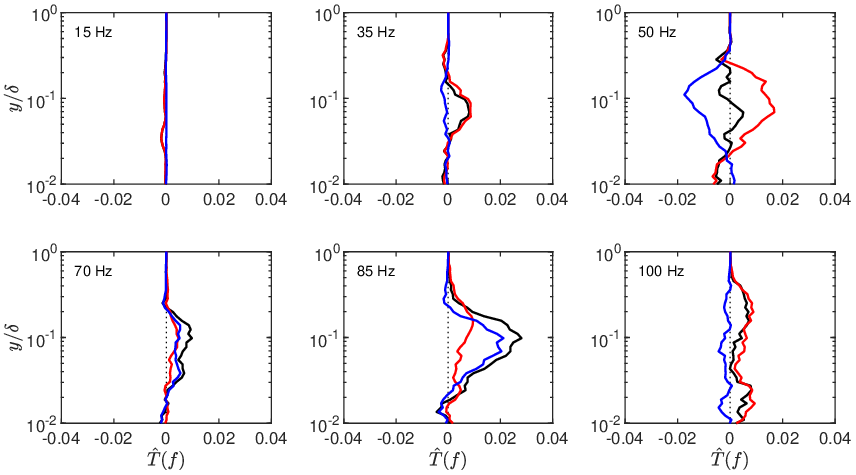}
    	\caption{Profiles of the contributions to the streamwise triadic transfer from Equations \ref{eq:SpecTransfer} and \ref{eq:SpecTransferParts}. Blue lines show the sum combinations $\hat{T}_S$, red lines the difference combination $\hat{T}_D$ (both from Equation \ref{eq:SpecTransferParts}), and black lines are the total $\hat{T}$ from Equation \ref{eq:SpecTransfer}. }
        \label{fig:FreqTransfer2F}
\end{figure}

To better see the energy transfer mechanisms associated with the forcing frequencies and their triads, Fig.~\ref{fig:FreqTransfer2F} plots the contributions to the spectral energy transfer for the two forcing frequencies, their sum and difference, and their first harmonic. It is apparent that the 35 Hz signal predominantly sends energy towards higher frequencies as $\hat{T}(f)$ is positive, but note that the majority of the energy transfer is occurring through the difference triads $\hat{T}_D$, which means the energy goes to $f_1 + 35  = f_2$ Hz.  Conversely, the 50 Hz signal has a mixture of summing and difference components, indicating that both forward and reverse cascade processes are taking place (which was also clearly indicated from the biphase profiles associated with 50 Hz in Fig.~\ref{fig:slices}).  The overall balance indicates that 50 Hz, while injected into the system from external forcing, ends up coupling with both higher and lower frequencies throughout the boundary layer. 
Additionally, it is clear that the greatest amount of spectral energy transfer is happening between the two forcing frequencies and their additive triad. The two harmonic modes have smaller magnitudes, and the difference mode is very weak and solely comprised of a reverse energy transfer where the coupling is from a difference mode of $f_1 - f_2 = 35$ Hz.

Tying these results to the phase analysis and skewness, the peak energy transfer processes occur in the 35, 50, 70, and 85 Hz signals at a location of $y/\delta \approx 0.1$, which corresponds to the location of $S_u = 0$ in Fig.~\ref{fig:SkewnessCalc}. The 35 and 50 Hz peak positive values in Fig.~\ref{fig:FreqTransfer2F} are associated with the difference mode, meaning that all frequencies that subtract to 35 and 50 Hz experience maximal energy transfer near $y/\delta = 0.1$. This matches the results from the average phase in Fig.~\ref{fig:bulkphase} in that the higher frequency region has an average phase of $\varphi = -\pi/2$ at the same wall-normal location, aligning with this result with the difference mode. These results align with both \citet{cui2021biphase} and \citet{jacobi2021interactions} in their studies of wall-bounded turbulent flows. Additionally, \citet{duvvuri2016nonlinear} found that the critical layer for these 6 frequencies exists around the region of $0.05\leq y/\delta \leq 0.08$ which coincides with the locations of maximum energy transfer seen in Fig.~\ref{fig:FreqTransfer2F}.

Recall that $\hat{T}(f)$ is only a measure of the streamwise contribution to the nonlinear energy transfer function, and other terms in the spanwise and wall-normal directions will also contribute to the transfer of energy between frequencies. As the perturbation is caused by a wall-normal oscillation of a plate across the spanwise direction, there are energy transfer mechanisms between the $u$, $v$, and $w$ components of velocity, in addition to wall-normal and spanwise directions of transfer, that are not captured in this streamwise velocity signal. The study of \citet{apostolidis2023turbulent} unveiled the nature of energy transfer across scales in the intermediate layer of a turbulent channel flow through integrating shells over all three physical dimensions, which is not possible with the limited data of a hotwire signal. However, coupling this measure of $\hat{T}(f)$ with the bicoherence and biphase profiles helps to highlight the energy transfer mechanisms that occur between the externally excited modes, even if a full picture of the energy cascade cannot be completely revealed.

\section{Conclusions}

The bispectrum has been shown to serve as a tool for spectrally decomposing and highlighting triadic interactions in turbulent velocity signals. The normalized bispectrum, the bicoherence, gives a measure of the relative amount of energy that exists at the frequency resulting from the phase alignment of two lower frequencies. The phase of the bispectrum, the biphase, reveals whether the higher frequency of the triadic interaction is leading or lagging the lower frequencies, providing insight into forward and reverse cascades of energy. A weighted summation of frequency pairs from the bispectrum gives both the skewness and asymmetry of the signal, and a pre-multiplied summation of imaginary components yields the interscale energy transfer for the streamwise velocity. Most notable is that these measures, including the nature of the forward or reverse cascade, are all derived from a direct calculation of the Fourier transform of the signal and do not require any filtering or arbitrary scale isolation. Utilizing these tools on the three boundary layer profiles, two of which are perturbed in different fashions, allow a clear insight into the way in which the triadic interactions are laid bare.

The magnitude of the bispectrum (Fig.~\ref{fig:bispec}) clearly shows the external perturbations in both the single- and two-frequency forcing cases, as well as making clear that the forcing frequencies form triads with all other frequencies in the flow. In the two-frequency case, the sum, difference, and harmonics of 35 and 50 Hz are clearly the highest magnitude locations. However, the magnitude of the bispectrum decreases quickly, so while the forcing frequencies and their triadic interactions are clear, the higher-frequencies regions will be negligibly small. The normalized measure, or bicoherence (Figs.~\ref{fig:bicohere}(a), (b), and \ref{fig:slices}(a)), provides both a way to see the effect of quadratic phase coupling across the spectral domain as well as assess the overall amount of energy at the coupled mode that is a direct result of the triadic interaction. The two-frequency forcing case had clear coupling of the forced modes from the VLSM range out to the highest frequencies. Additionally, as all turbulent signals will have some aspect of phase coupling due to the nature of the nonlinearity in the Navier-Stokes equations, the signal can be masked for statistical significance to show the strongest triadic interactions. Once again, the two-frequency forcing case has clear indications of triads not only with the forcing frequencies, but triads of the triadic interactions start to become unveiled with combinations such as $15+70 = 85$ Hz and $15 + 85 = 100$ Hz. 

The phase of the bispectrum is a measure of the phase angle between the high frequency (small scale) and the two lower frequencies (large scales), which is directly related to the energy cascade \cite{cui2021biphase,jacobi2021interactions}. The phase relationship within an individual triad at a given wall-normal position is directly revealed with $\beta(f_1,f_2)$ in Fig.~\ref{fig:bicohere}(c), but is better assessed by fixing one of the lower frequencies and looking through the entire boundary layer, as seen in Fig.~\ref{fig:slices}(b). This study shows that the forcing modes and their triadic interactions all have a reverse cascade in the near-wall region. Whether this is the result of forcing frequencies in the large-scale regions of $f<f_0 = 212$ Hz for the outer scale \cite[see for instance][]{marusic2021energy}, the alignment of fluctuation scales to cause local vortex stretching \cite[details in][]{apostolidis2023turbulent}, or some other mechanism warrants further investigation.

Summing the bispectrum over triadically consistent frequencies reveals the spectral contributions to the velocity skewness and asymmetry. These measures make clear that any nonzero skewness is an indication of triadic interactions, although phase-leading and phase-lagging scale contributions could sum to zero as well. The external forcing works to decrease the skewness, making it cross zero closer to the wall. All forcing frequencies and their triads contributed negative values to $S_u(f)$ throughout the boundary layer. In contrast, the asymmetry increases with forcing and has a wider range of influence throughout the boundary layer. The forcing frequencies themselves contributed positive values to $A_u(f)$ throughout the boundary layer, while the triadic interactions were smaller in magnitude and negative. As they asymmetry is related to the slope of the signal, there is a possibility that this could be another measure that relates to sweep and ejection events, although concurrent measures of the streamwise and wall-normal velocity would be needed to confirm this.

Taking the phase of the spectral breakdown of skewness and asymmetry provides an average measure $\varphi(f)$ of all triadic interactions that contribute to a higher frequency. This resulted in two significant findings: first, small scales appear to be in phase with large scales near the wall and increase in their phase lead further from the wall, consistent with prior studies; second, that the location of $\varphi = -\pi/2$ moves closer to the wall with added forcing, indicating the location of maximal energy transfer is pulled towards the location of the critical layer associated with the forcing modes. Another interesting finding is that all triads below the outer scale of $f_o=212$ Hz have a mixture of phase lead and phase lag, while the average phase beyond this scale is always negative, indicating the forward cascade. Analysis across a range of flow Reynolds numbers and physical scales is necessary to determine if this relationship is universal or simply a coincidence.
  
Finally, the imaginary component of the bispectrum can be summed in triadically consistent combinations of wavenumbers to produce the spectral energy transfer function in the streamwise direction. The contributions to this spectral energy transfer from additive triads and difference triads are consistent with the findings in the bicoherence, biphase, skewness, and average phase. The location of maximal energy transfer for these triads coincides with the location of $\varphi = -\pi/2$, further highlighting the usefulness of the average phase calculation. 

These findings illustrate the usefulness of the bispectrum and associated normalizations and measures when it comes to assessing the nature of energy transfer and triadic interactions in turbulence, all without any need for scale isolation or artificial filtering of the signal. By analyzing a canonical boundary layer that has clear perturbation signals applied in the near-wall region, the triadic interactions and the bispectrum's relation to them become easy to interpret. With the findings of recent studies on the application of external forcing frequencies to lower wall shear \cite{marusic2021energy} and the resulting interactions of the forcing mode with the inner and outer scales \cite{deshpande2023relationship}, these analysis techniques may provide additional information on the nature of how external forcing disrupts and alters turbulent flow. The limitation in this study of only investigating the streamwise component of velocity means that the full nature of the energy cascade cannot be revealed, but extending this technique to multiple components of velocity and pressure fluctuations should serve as a compliment to existing and future studies.

\section*{Acknowledgments}

The data were originally generated with support from AFOSR (award no. FA 9550-12-1-0469, PI B. McKeon, program manager D. Smith).


\bibliography{references}

@article{adrian2007hairpin,
  title={Hairpin vortex organization in wall turbulence},
  author={Adrian, Ronald J},
  journal={Physics of Fluids},
  volume={19},
  number={4},
  pages = {041301},
  year={2007},
  publisher={AIP Publishing}
}

@article{alexakis2018cascades,
  title={Cascades and transitions in turbulent flows},
  author={Alexakis, Alexandros and Biferale, Luca},
  journal={Physics Reports},
  volume={767},
  pages={1--101},
  year={2018},
  publisher={Elsevier}
}

@article{apostolidis2023turbulent,
  title={Turbulent cascade in fully developed turbulent channel flow},
  author={Apostolidis, Argyrios and Laval, Jean-Philippe and Vassilicos, JC},
  journal={Journal of Fluid Mechanics},
  volume={967},
  pages={A22},
  year={2023}
}

@article{baars2015wavelet,
  title={Wavelet analysis of wall turbulence to study large-scale modulation of small scales},
  author={Baars, Woutijn J and Talluru, KM and Hutchins, Nicholas and Marusic, Ivan},
  journal={Experiments in Fluids},
  volume={56},
  number={10},
  pages={188},
  year={2015},
  publisher={Springer}
}

@article{bandyopadhyay1984coupling,
  title={The coupling between scales in shear flows},
  author={Bandyopadhyay, Promode R and Hussain, AKMF},
  journal={Physics of Fluids},
  volume={27},
  number={9},
  pages={2221--2228},
  year={1984},
  publisher={AIP Publishing}
}

@article{brown1977large,
  title={Large structure in a turbulent boundary layer},
  author={Brown, Garry L and Thomas, Andrew SW},
  journal={Physics of Fluids},
  volume={20},
  number={10},
  pages={S243--S252},
  year={1977},
  publisher={AIP Publishing}
}

@article{chokani2005nonlinear,
  title={Nonlinear evolution of Mack modes in a hypersonic boundary layer},
  author={Chokani, Ndaona},
  journal={Physics of Fluids},
  volume={17},
  number={1},
  pages={014102},
  year={2005},
  publisher={American Institute of Physics}
}

@article{chung2010large,
  title={Large-eddy simulation of large-scale structures in long channel flow},
  author={Chung, Daniel and Mckeon, Beverley J},
  journal={Journal of Fluid Mechanics},
  volume={661},
  pages={341--364},
  year={2010},
  publisher={Cambridge University Press}
}

@article{cui2021biphase,
  title={Biphase as a diagnostic for scale interactions in wall-bounded turbulence},
  author={Cui, G and Jacobi, I},
  journal={Physical Review Fluids},
  volume={6},
  number={1},
  pages={014604},
  year={2021},
  publisher={APS}
}

@article{deshpande2023relationship,
  title={On the relationship between manipulated inter-scale phase and energy-efficient turbulent drag reduction},
  author={Deshpande, Rahul and Chandran, Dileep and Smits, Alexander J and Marusic, Ivan},
  journal={Journal of Fluid Mechanics},
  volume={972},
  pages={A12},
  year={2023},
  publisher={Cambridge University Press}
}

@inproceedings{duvvuri2014phase,
  title={Phase relationships in presence of a synthetic large-scale in a turbulent boundary layer},
  author={Duvvuri, Subrahmanyam and McKeon, Beverley J},
  booktitle={44th AIAA fluid dynamics conference},
  pages={2883},
  year={2014}
}

@article{duvvuri2015triadic,
  title={Triadic scale interactions in a turbulent boundary layer},
  author={Duvvuri, Subrahmanyam and McKeon, Beverley J},
  journal={Journal of Fluid Mechanics},
  volume={767},
  pages={R4},
  year={2015},
  publisher={Cambridge University Press}
}

@article{duvvuri2016nonlinear,
  title={Nonlinear interactions isolated through scale synthesis in experimental wall turbulence},
  author={Duvvuri, Subrahmanyam and McKeon, Beverley},
  journal={Physical Review Fluids},
  volume={1},
  number={3},
  pages={032401},
  year={2016},
  publisher={APS}
}

@article{duvvuri2017phase,
  title={Phase relations in a forced turbulent boundary layer: implications for modelling of high Reynolds number wall turbulence},
  author={Duvvuri, Subrahmanyam and McKeon, Beverley},
  journal={Philosophical Transactions of the Royal Society A: Mathematical, Physical and Engineering Sciences},
  volume={375},
  number={2089},
  pages={20160080},
  year={2017},
  publisher={The Royal Society Publishing}
}

@article{elgar1985observations,
  title={Observations of bispectra of shoaling surface gravity waves},
  author={Elgar, Steve and Guza, RT},
  journal={Journal of Fluid Mechanics},
  volume={161},
  pages={425--448},
  year={1985},
  publisher={Cambridge University Press}
}

@article{elgar1987relationships,
  title={Relationships involving third moments and bispectra of a harmonic process},
  author={Elgar, Steve},
  journal={IEEE Transactions on Acoustics, Speech, and Signal Processing},
  volume={35},
  number={12},
  pages={1725--1726},
  year={1987},
  publisher={IEEE}
}

@article{elgar1989statistics,
  title={Statistics of bicoherence and biphase},
  author={Elgar, Steve and Sebert, Gloria},
  journal={Journal of Geophysical Research: Oceans},
  volume={94},
  number={C8},
  pages={10993--10998},
  year={1989},
  publisher={Wiley Online Library}
}

@inproceedings{fackrell1995quadratic,
   author = {J.W.A. Fackrell and S. McLaughlin},
   title = {Quadratic phase coupling detection using higher order statistics},
   journal = {IET Conference Proceedings},
   booktitle={IEE Colloquium on `Higher Order Statistics in Signal Processing: Are They of Any Use?'},   
   year = {1995},
   month = {January},
   pages = {9-9(1)},
   publisher ={Institution of Engineering and Technology}
}

@phdthesis{fackrell1996bispectral,
    author = {Fackrell, Justin WA},
    title={Bispectral analysis of speech signals},
    school = {University of Edinburgh},
    year = {1997}
}

@article{george2017detecting,
  title={Detecting dynamical states from noisy time series using bicoherence},
  author={George, Sandip V and Ambika, G and Misra, R},
  journal={Nonlinear Dynamics},
  volume={89},
  number={1},
  pages={465--479},
  year={2017},
  publisher={Springer}
}

@article{haubrich1965earth,
  title={Earth noise, 5 to 500 millicycles per second: 1. Spectral stationarity, normality, and nonlinearity},
  author={Haubrich, Richard A},
  journal={Journal of Geophysical Research},
  volume={70},
  number={6},
  pages={1415--1427},
  year={1965},
  publisher={Wiley Online Library}
}

@article{he2017space,
  title={Space-time correlations and dynamic coupling in turbulent flows},
  author={He, Guowei and Jin, Guodong and Yang, Yue},
  journal={Annual Review of Fluid Mechanics},
  volume={49},
  number={1},
  pages={51--70},
  year={2017},
  publisher={Annual Reviews}
}

@article{HINICH2005Normalizing,
title = {Normalizing bispectra},
journal = {Journal of Statistical Planning and Inference},
volume = {130},
number = {1},
pages = {405-411},
year = {2005},
issn = {0378-3758},
doi = {https://doi.org/10.1016/j.jspi.2003.12.022},
url = {https://www.sciencedirect.com/science/article/pii/S0378375804002745},
author = {Melvin J. Hinich and Murray Wolinsky}
}

@article{hutchins2007large,
  title={Large-scale influences in near-wall turbulence},
  author={Hutchins, Nicholas and Marusic, Ivan},
  journal={Philosophical Transactions of the Royal Society A: Mathematical, Physical and Engineering Sciences},
  volume={365},
  number={1852},
  pages={647--664},
  year={2007},
  publisher={The Royal Society London}
}

@article{hutchins2011three,
  title={Three-dimensional conditional structure of a high-Reynolds-number turbulent boundary layer},
  author={Hutchins, Nicholas and Monty, Jason P and Ganapathisubramani, Bharathram and Ng, Henry Chi-Hin and Marusic, Ivan},
  journal={Journal of Fluid Mechanics},
  volume={673},
  pages={255--285},
  year={2011},
  publisher={Cambridge University Press}
}

@article{jacobi2013phase,
  title={Phase relationships between large and small scales in the turbulent boundary layer},
  author={Jacobi, I and McKeon, BJ},
  journal={Experiments in Fluids},
  volume={54},
  number={3},
  pages={1481},
  year={2013},
  publisher={Springer}
}

@article{jacobi2021interactions,
  title={Interactions between scales in wall turbulence: phase relationships, amplitude modulation and the importance of critical layers},
  author={Jacobi, Ian and Chung, Daniel and Duvvuri, Subrahmanyam and McKeon, Beverley J},
  journal={Journal of Fluid Mechanics},
  volume={914},
  pages={A7},
  year={2021},
  publisher={Cambridge University Press}
}

@article{jamvsek2003time,
  title={Time-phase bispectral analysis},
  author={Jam{\v{s}}ek, Janez and Stefanovska, Aneta and McClintock, Peter VE and Khovanov, Igor A},
  journal={Physical Review E},
  volume={68},
  number={1},
  pages={016201},
  year={2003},
  publisher={APS}
}

@article{jeffries1998experience,
  title={Experience with bicoherence of electrical power for condition monitoring of wind turbine blades},
  author={Jeffries, WQ and Chambers, Jonathan A and Infield, David G},
  journal={IEE Proceedings - Vision, Image and Signal Processing},
  volume={145},
  number={3},
  pages={141--148},
  year={1998},
  publisher={IET}
}

@article{jimenez2012cascades,
  title={Cascades in wall-bounded turbulence},
  author={Jim{\'e}nez, Javier},
  journal={Annual Review of Fluid Mechanics},
  volume={44},
  number={1},
  pages={27--45},
  year={2012},
  publisher={Annual Reviews}
}

@article{katul2006relative,
  title={The relative importance of ejections and sweeps to momentum transfer in the atmospheric boundary layer},
  author={Katul, Gabriel and Poggi, Davide and Cava, Daniela and Finnigan, John},
  journal={Boundary-Layer Meteorology},
  volume={120},
  number={3},
  pages={367--375},
  year={2006},
  publisher={Springer}
}

@article{kim1979digital,
  title={Digital bispectral analysis and its applications to nonlinear wave interactions},
  author={Kim, Young C and Powers, Edward J},
  journal={IEEE Transactions on Plasma Science},
  volume={7},
  number={2},
  pages={120--131},
  year={1979},
  publisher={IEEE}
}

@article{kline1967structure,
  title={The structure of turbulent boundary layers},
  author={Kline, Stephen J and Reynolds, William C and Schraub, Frederic Anthony and Runstadler, Peter W},
  journal={Journal of Fluid Mechanics},
  volume={30},
  number={4},
  pages={741--773},
  year={1967},
  publisher={Cambridge University Press}
}

@article{kovach2018bispectrum,
  title={The bispectrum and its relationship to phase-amplitude coupling},
  author={Kovach, Christopher K and Oya, Hiroyuki and Kawasaki, Hiroto},
  journal={Neuroimage},
  volume={173},
  pages={518--539},
  year={2018},
  publisher={Elsevier}
}

@article{lee2019spectral,
  title={Spectral analysis of the budget equation in turbulent channel flows at high Reynolds number},
  author={Lee, Myoungkyu and Moser, Robert D},
  journal={Journal of Fluid Mechanics},
  volume={860},
  pages={886--938},
  year={2019},
  publisher={Cambridge University Press}
}

@article{Li2023Quantifying,
  title = {Quantifying inner-outer interactions in noncanonical wall-bounded flows},
  author = {Li, Mogeng and Baars, Woutijn J. and Marusic, Ivan and Hutchins, Nicholas},
  journal = {Physical Review Fluids},
  volume = {8},
  issue = {8},
  pages = {084602},
  numpages = {19},
  year = {2023},
  month = {Aug},
  publisher = {American Physical Society},
}

@article{liu2022evolution,
  title={Evolution of turbulent kinetic energy during the entire sandstorm process},
  author={Liu, Hongyou and Shi, Yanxiong and Zheng, Xiaojing},
  journal={Atmospheric Chemistry and Physics},
  volume={22},
  number={13},
  pages={8787--8803},
  year={2022},
  publisher={Copernicus GmbH}
}

@article{lozier2024revisiting,
  title={Revisiting amplitude modulation in non-canonical wall-turbulence through high-Reynolds number experimental data},
  author={Lozier, Mitchell and Marusic, Ivan and Deshpande, Rahul},
  journal={Physical Review Fluids},
  volume={9},
  number={12},
  pages={124602},
  year={2024},
  publisher={APS}
}

@inproceedings{lyu2020correlation,
  title={Correlation analysis between forced oscillation modes caused by wind power},
  author={Lyu, Wan and Gu, Wen and Wu, Xi and Yang, Hongyu and Jiang, Chen and Xu, Shanshan},
  booktitle={Journal of Physics: Conference Series},
  volume={1639},
  number={1},
  pages={012062},
  year={2020},
  organization={IOP Publishing}
}

@article{marusic2021energy,
  title={An energy-efficient pathway to turbulent drag reduction},
  author={Marusic, Ivan and Chandran, Dileep and Rouhi, Amirreza and Fu, Matt K and Wine, David and Holloway, Brian and Chung, Daniel and Smits, Alexander J},
  journal={Nature Communications},
  volume={12},
  number={1},
  pages={5805},
  year={2021},
  publisher={Nature Publishing Group UK London}
}

@article{mathis2009large,
  title={Large-scale amplitude modulation of the small-scale structures in turbulent boundary layers},
  author={Mathis, Romain and Hutchins, Nicholas and Marusic, Ivan},
  journal={Journal of Fluid Mechanics},
  volume={628},
  pages={311--337},
  year={2009},
  publisher={Cambridge University Press}
}

@article{matsuoka1984phase,
  title={Phase estimation using the bispectrum},
  author={Matsuoka, Toshifumi and Ulrych, Tad J},
  journal={Proceedings of the IEEE},
  volume={72},
  number={10},
  pages={1403--1411},
  year={1984},
  publisher={IEEE}
}

@inproceedings{midya2022spectral,
  title={ON THE SPECTRAL DECOMPOSITION OF SKEWNESS IN CANONCIAL AND ACTUATED TURBULENT BOUNDARY LAYERS},
  author={Midya, S and Thomas, FO and Gordeyev, SV},
  booktitle={Proceedings of the 12th International Symposium on Turbulence and Shear Flow Phenomena (TSFP12), Osaka, Japan},
  year={2022}
}

@article{ning1989bispectral,
  title={Bispectral analysis of the rat EEG during various vigilance states},
  author={Ning, Taikang and Bronzino, Joseph D},
  journal={IEEE Transactions on Biomedical Engineering},
  volume={36},
  number={4},
  pages={497--499},
  year={1989},
  publisher={IEEE}
}

@article{pham2020autism,
  title={Autism spectrum disorder diagnostic system using HOS bispectrum with EEG signals},
  author={Pham, The-Hanh and Vicnesh, Jahmunah and Wei, Joel Koh En and Oh, Shu Lih and Arunkumar, N and Abdulhay, Enas W and Ciaccio, Edward J and Acharya, U Rajendra},
  journal={International Journal of Environmental Research and Public Health},
  volume={17},
  number={3},
  pages={971},
  year={2020},
  publisher={MDPI}
}

@book{Pope2000, place={Cambridge}, title={Turbulent Flows}, publisher={Cambridge University Press}, author={Pope, Stephen B.}, year={2000}}

@article{quadrio2011drag,
  title={Drag reduction in turbulent boundary layers by in-plane wall motion},
  author={Quadrio, Maurizio},
  journal={Philosophical Transactions of the Royal Society A: Mathematical, Physical and Engineering Sciences},
  volume={369},
  number={1940},
  pages={1428--1442},
  year={2011},
  publisher={The Royal Society Publishing}
}

@article{rao1971bursting,
  title={The `bursting' phenomenon in a turbulent boundary layer},
  author={Rao, K. N. and Narasimha, R. and Narayanan, M. A. B.},
  journal={Journal of Fluid Mechanics},
  volume={48},
  pages={339--352},
  year={1971},
  publisher={Cambridge University Press}
}

@article{ricco2021review,
  title={A review of turbulent skin-friction drag reduction by near-wall transverse forcing},
  author={Ricco, Pierre and Skote, Martin and Leschziner, Michael A},
  journal={Progress in Aerospace Sciences},
  volume={123},
  pages={100713},
  year={2021},
  publisher={Elsevier}
}

@article{ruessink2009modeling,
author = {Ruessink, B. G. and van den Berg, T. J. J. and van Rijn, L. C.},
title = {Modeling sediment transport beneath skewed asymmetric waves above a plane bed},
journal = {Journal of Geophysical Research: Oceans},
volume = {114},
number = {C11},
pages = {C11021},
year = {2009}
}

@article{saxton2022amplitude,
  title={Amplitude and wall-normal distance variation of small scales in turbulent boundary layers},
  author={Saxton-Fox, Theresa and Lozano-Dur{\'a}n, Adri{\'a}n and McKeon, Beverley J},
  journal={Physical Review Fluids},
  volume={7},
  number={1},
  pages={014606},
  year={2022},
  publisher={APS}
}

@article{shils1996bispectral,
  title={Bispectral analysis of visual interactions in humans},
  author={Shils, JL and Litt, M and Skolnick, BE and Stecker, MM},
  journal={Electroencephalography and Clinical Neurophysiology},
  volume={98},
  number={2},
  pages={113--125},
  year={1996},
  publisher={Elsevier}
}

@article{sigl1994introduction,
  title={An introduction to bispectral analysis for the electroencephalogram},
  author={Sigl, Jeffrey C and Chamoun, Nassib G},
  journal={Journal of Clinical Monitoring},
  volume={10},
  number={6},
  pages={392--404},
  year={1994},
  publisher={Springer}
}

@article{slomka2018nature,
  title={The nature of triad interactions in active turbulence},
  author={S{\l}omka, Jonasz and Suwara, Piotr and Dunkel, J{\"o}rn},
  journal={Journal of Fluid Mechanics},
  volume={841},
  pages={702--731},
  year={2018},
  publisher={Cambridge University Press}
}

@article{wang2021coherent,
  title={Coherent structures associated with interscale energy transfer in turbulent channel flows},
  author={Wang, Hongping and Yang, Zixuan and Wu, Ting and Wang, Shizhao},
  journal={Physical Review Fluids},
  volume={6},
  number={10},
  pages={104601},
  year={2021},
  publisher={APS}
}

@article{wang2024role,
  title={Role of Fourier phase dynamics in decaying turbulence},
  author={Wang, Chuhan and Fang, Le and Wang, Zhan and Xu, Chunxiao},
  journal={Physical Review Fluids},
  volume={9},
  number={11},
  pages={114603},
  year={2024},
  publisher={APS}
}

@article{xiao2009physical,
  title={Physical mechanism of the inverse energy cascade of two-dimensional turbulence: a numerical investigation},
  author={Xiao, Z and Wan, Minping and Chen, Shiyi and Eyink, Gregory L},
  journal={Journal of Fluid Mechanics},
  volume={619},
  pages={1--44},
  year={2009},
  publisher={Cambridge University Press}
}

@ARTICLE{xie2019effect,
  author={Xie, Dengfeng and Chen, Kun-Shan and Yang, Xiaofeng},
  journal={IEEE Journal of Selected Topics in Applied Earth Observations and Remote Sensing}, 
  title={Effect of Bispectrum on Radar Backscattering From Non-Gaussian Sea Surface}, 
  year={2019},
  volume={12},
  number={11},
  pages={4367-4378},
  doi={10.1109/JSTARS.2019.2946934}}

@article{yagiz2012drag,
  title={Drag minimization using active and passive flow control techniques},
  author={Yagiz, Bedri and Kandil, Osama and Pehlivanoglu, Y Volkan},
  journal={Aerospace Science and Technology},
  volume={17},
  number={1},
  pages={21--31},
  year={2012},
  publisher={Elsevier}
}

@article{zandvoort2021defining,
  title={Defining the filter parameters for phase-amplitude coupling from a bispectral point of view},
  author={Zandvoort, Coen S and Nolte, Guido},
  journal={Journal of Neuroscience Methods},
  volume={350},
  pages={109032},
  year={2021},
  publisher={Elsevier}
}

@article{zhang2020active,
  title={Active control for wall drag reduction: Methods, mechanisms and performance},
  author={Zhang, Lu and Shan, Xiaobiao and Xie, Tao},
  journal={IEEE Access},
  volume={8},
  pages={7039--7057},
  year={2020},
  publisher={IEEE}
}

\end{document}